\documentstyle[12pt,epsf]{article}
\def\slash#1{#1\hskip-6pt/\hskip6pt}
\def\nicefrac#1#2{\hbox{${#1\over #2}$}}

\begin{document}

\begin{titlepage}

\setcounter{page}{1}

\rightline{JLAB--THY-97-39}
\rightline{UFIFT--HEP-97-25}

\vfill
\begin{center}
 {\Large \bf Transverse Spin Dependent Drell Yan in QCD }
\centerline{\large \bf to $O(\alpha_s^2)$  at Large $p_T$ (I).} 
\centerline{\bf Virtual Corrections and Methods for the Real Emissions}

\vfill
\vfill
 {\large Sanghyeon Chang$^{*}$\footnote{
        E-mail address: schang@phys.ufl.edu},
        Claudio Corian\`{o}$^{**}$\footnote{
        E-mail address: coriano@jlabs2.jlab.org}   
        and John K. Elwood$^{* \dagger}$\footnote{
        E-mail address: elwoodjo@tusc.kent.edu}}
\\
\vspace{.12in}
 {\it $^{*}$   Institute for Fundamental Theory, Department of Physics, \\
        University of Florida, Gainesville, FL 32611, 
        USA}
\\
\vspace{.12in}
{\it $^{**}$ Theory Group, Jefferson Lab, Newport News, VA 23606, USA}
\\
\vspace{.12in}
{\it $^{\dagger}$ Department of Physics, Kent State University, Kent, OH 44242, USA}
\end{center}
\vfill
\begin{abstract}

We investigate the role of the transverse spin dependence in Drell Yan 
lepton pair production to NLO in QCD at parton level.
In our analysis we deal with the large $p_T$ distributions.
We give very compact expressions for the virtual $O(\alpha_s^2)$ 
corrections to the cross section and show that the singularities 
factorize. The study is performed in the $\overline{MS}$ 
scheme in Dimensional Regularization, and with the t'Hooft-Veltman prescription for $\gamma_5$. A discussion of the structure of the real 
emissions is included, and detailed methods for the study of these 
contributions are formulated.

\end{abstract}

\end{titlepage}

\setcounter{footnote}{0}

% ========================= DEFINITIONS ===================================
\def\beq{\begin{equation}}
\def\eeq{\end{equation}}
\def\beqn{\begin{eqnarray}}
\def\eeqn{\end{eqnarray}}

\def\ie{{\it i.e.}}
\def\eg{{\it e.g.}}
\def\half{{\textstyle{1\over 2}}}
\def\third{{\textstyle {1\over3}}}
\def\quarter{{\textstyle {1\over4}}}
\def\m{{\tt -}}

\def\p{{\tt +}}

\def\slash#1{#1\hskip-6pt/\hskip6pt}
\def\slasher#1{#1\hskip-5pt/}
\def\slk{\slash{k}}
\def\GeV{\,{\rm GeV}}
\def\TeV{\,{\rm TeV}}
\def\y{\,{\rm y}}

\def\l{\langle}
\def\r{\rangle}

\setcounter{footnote}{0}
\newcommand{\beqa}{\begin{eqnarray}}
\newcommand{\eeqa}{\end{eqnarray}}
\newcommand{\eps}{\epsilon}

\section{Introduction}

There has been considerable recent interest in the study of 
polarized DIS collisions at HERA, both from the theoretical and the 
experimental side. Most of the work, so far, has been directed 
toward the analysis of the (longitudinal) spin 
content of the proton, in connection with the so called ``spin crisis'' and 
the puzzling results of the EMC collaboration (see \cite{ramsey,Cheng}). 
The next step of this ongoing analysis is to study, for various ranges of 
$x$ and $Q^2$, the leading twist  (polarized) 
parton distributions in p-p collisions in a different experimental setting.
Such a program will be possible in the not distant future at polarized hadron colliders 
such as RHIC, using polarized p-p incoming states. 

A direct gluon coupling, absent in DIS, will allow us to study the gluon 
content of the nucleons in an independent manner. The coupling is 
non-anomalous and, therefore, all the theoretical 
ambiguities associated with the 
scale-independence of the axial anomaly are not present.

There is also much more to be gained from these studies. 

It is well known that in polarized DIS, the leading twist 
transverse spin distribution $h_1(x)$ decouples, being chirally odd, 
and one possible way to get some information about its parton content is to 
resort to polarized p-p collisions.

\begin{figure}[h]
\centerline{\epsfbox{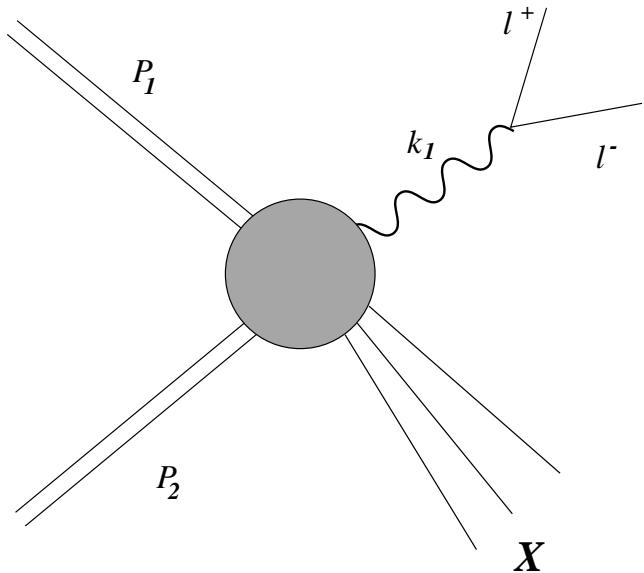}}
\caption{The Drell-Yan process}
\label{cdy}
\end{figure}

There are at least two reasons to justify a detailed NLO study of 
the Drell Yan (DY) process,
Fig.~\ref{cdy}, both in the case of the longitudinal and of the 
transverse asymmetries. The first one is that the final state is particularly ``clean'', 
due to the presence of a lepton pair.  
This means that the photon fragmentation contribution 
does not affect the momentum distribution of the 
final state. In particular,
 the scale dependence associated 
to a fragmenting photon is absent. 

The second reason is that transverse asymmetries 
are sensitive to gluon distributions only through radiative corrections. 
In previous works, the complete (real plus virtual) 
NLO corrections to the non singlet DY 
cross section with longitudinally polarized initial states have been 
presented (for the nonzero $p_T\equiv q_T$ distributions). 
In this work we make the first step toward extending 
these calculations to the case of transversely polarized incoming states. 

Although much of the experimental data are usually collected at small $p_T$, 
the study of the distributions at large $p_T$ remains, nevertheless, 
a challenging and complex problem.

As a results of these studies, we are able to 
perform nontrivial perturbative checks of factorization in QCD and we 
developed more complete strategies for the investigation of spin dependent cross sections. Most of the techniques that we consider in this and in former 
related 
\cite{CCFG} work are, in fact, applicable to the easier case of polarized $e^+ e^-$ collisions.

Our analysis is organized as follows. 
After a discussion of the main features of the various cross sections in the the study of the Drell Yan process, we 
analyze the properties of the cross section $d\sigma/d^4 q \,\,d\Omega$. 
 
We separate the contributions to this cross section into ``diagonal'' and 
``non-diagonal'' terms, and then we give very compact results for 
the virtual corrections to the ``diagonal'' terms. 
The term ``diagonal'' refers to those contributions obtained by contracting 
the hadronic tensor (at parton level) with the $g^{\mu\nu}$ part of the leptonic tensor, while the ``non-diagonal'' part contains the momenta of the lepton pair (here denoted as $v_1^\mu v_2^\nu$). 
We show that these ``diagonal'' virtual corrections factorize. 
The result is presented in a remarkably simple form. 
The reason for performing this separation comes from the structure 
of the factorized cross section in terms of parton distributions as given originally by Ralston and Soper in the parton model \cite{RS}. We will come back to this point briefly in one of the following sections.

Then we move to a detailed 
discussion of the real emission contributions and show that in our cross section the integration over 3 of the five final-state partons is 
sufficient to cancel the singularities of the virtual corrections.  
We will not give here simplified results for the real emissions, which will 
be presented elsewhere, but we set up the stage for a complete analysis 
of the real emissions to NLO. Some relevant technical aspects of the calculation have been included in various 
appendices. A very efficient method to perform the renormalization of all the diagrams -which is of simple symbolic implementation- is illustrated in some detail, since it is of general applicability. In a concluding section 
we summarize our analysis and elaborate on future developments 
and unsolved issues of the theory.

\begin{figure}[h]
\centerline{\epsfbox{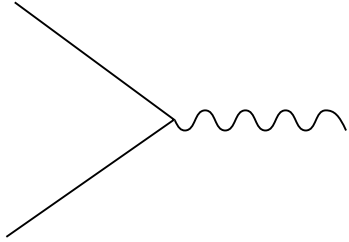}}
\caption{Leading order contribution to the Drell-Yan process}
\label{modlow}
\end{figure}

\section{Zero $p_T$ distributions. General Features}

Let $H_1$ and $H_2$ be two polarized hadron that collide 
thereby producing a heavy photon and some hadronic remnants (X)

\beq
H_1 + H_2\to \gamma^* + X. 
\eeq

At parton level the process goes through the elementary scattering 
\beq
a + b \to \gamma^* \to l^+ + l^-,
\eeq
and the leading channel is the annihilation channel. 
The process is purely electromagnetic at the lowest order. 
In fact, Fig.~\ref{modlow} is the leading order contribution 
to the DY process. 
At this order, the lepton pair is produced with a total 
zero $q_T$, i.e. back-to-back.

If we include radiative corrections -with an additional parton in the final 
state beside the photon-
then the lepton pair gets a 
nonzero $q_T$. Two cross sections are studied in general: $\sigma(Q^2)$ and $d\sigma/dy\, d x_F$, 
with y being the rapidity of the photon and $x_F$ is its ``Feynman's x''. 
Both cross sections are sufficiently inclusive and well defined in perturbation theory so that an 
analytic calculation is possible. 

\begin{figure}[h]
\epsfxsize=140mm
\centerline{\epsfbox{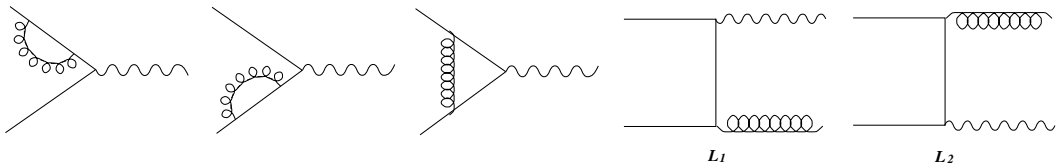}}
\caption{$\alpha_s$ order contributions}
\label{low}
\end{figure}

The NLO corrections to the $\sigma(Q^2)$ cross section are quite straightforward 
to obtain. At this order, for instance, the real emissions involve only 2 partons 
in the final state, while the virtual require the 1-loop renormalized 
$q\,\,\bar{q}\,\,\gamma^*$ vertex plus quark self-energy insertions 
(see Fig.~\ref{low}).

In this cross section one of the final state partons is integrated over 
(the gluon or the quark) and the transverse momentum dependence of the lepton pair 
is not completely resolved. 

A better accuracy is obtained 
 by going to NNLO, i.e. at order $\alpha_s^2$. 

For instance, in the non-singlet sector, all the diagrams $V_1,...,V_{11}$ 
(Fig.~7) are needed, plus there are contributions from the so called ``G'', ``F'' and ``H'' 
sets (see Figs.~4-6). Basically, the G-set involves real emissions of the form $q\bar{q}\to g g \gamma^*$; 
the set F involves both annihilation ($q\bar{q}\to q\bar{q} \gamma^*$) and scattering 
$(q\bar{q}\to q\bar{q} \gamma^*)$ diagrams, while the set $H$ involves $q q'\to q q' \gamma^*$ scattering diagrams. 

Additionally, one needs to evaluate all the corrections 
coming from the interference between the 2-loop (polarized) on-shell quark form factor 
and the lowest order process of Fig~\ref{modlow}. The latter are the only two-loop contributions needed for a 
complete NNLO calculation of the invariant mass distribution.  

In a completely anticommuting scheme for $\gamma_5$ this can be obtained directly from the literature, and differs from the result of refs. 
\cite{formfac} just by an overall helicity factor $(1-h)$. In the 
t'Hooft-Veltman scheme it is expected that additional helicity-violating terms will appear, since the on-shell form factor is not infrared (IR) safe. 
However, a redefinition of the polarized splitting function of the form 
$\Delta P_{qq}=P_{qq} + 4 C_F \epsilon (z-1) + O(\epsilon^2)$ 
should be sufficient to eliminate the helicity-violating contributions. 
However, as we have mentioned before, the invariant mass distribution has to be analyzed both in its diagonal and non-diagonal contributions in order to 
obtain information on the perturbative corrections to the transverse spin cross section. In fact, as we will discuss below, it is the 
structure of the parton model result for $\sigma^{TT}$ that forces us to 
consider a more general approach to the DY process, and renders the calculation far more complex 
than in the unpolarized or in the longitudinally polarized case.

\section{Nonzero $p_T$ Distributions}
Since there is a lot of information in the transverse momentum distribution of the lepton pair and 
since, experimentally, one needs to set cuts below a certain  $p_T$, 
the 
study of the transverse momentum distributions turns out to be of 
considerable interest. Various cross sections are studied for this purpose.
One of the most studied cross section is defined as $d\sigma/d^2Q_T$, or, equivalently as $d\sigma/dt\,\, du$. In the analysis of the virtual corrections 
we will concentrate on the latter. We remind that the diagonal part of the 
more general cross section $d\sigma/d^4q\, d\Omega$, with $\Omega$ defined 
to be the 
solid angle in the rest frame of the photon, is easily extracted from 
$d\sigma/dt\,\, du$. In their non-diagonal contributions the two cross sections are, of course, different. 

\begin{figure}[bh]
\centerline{\epsfbox{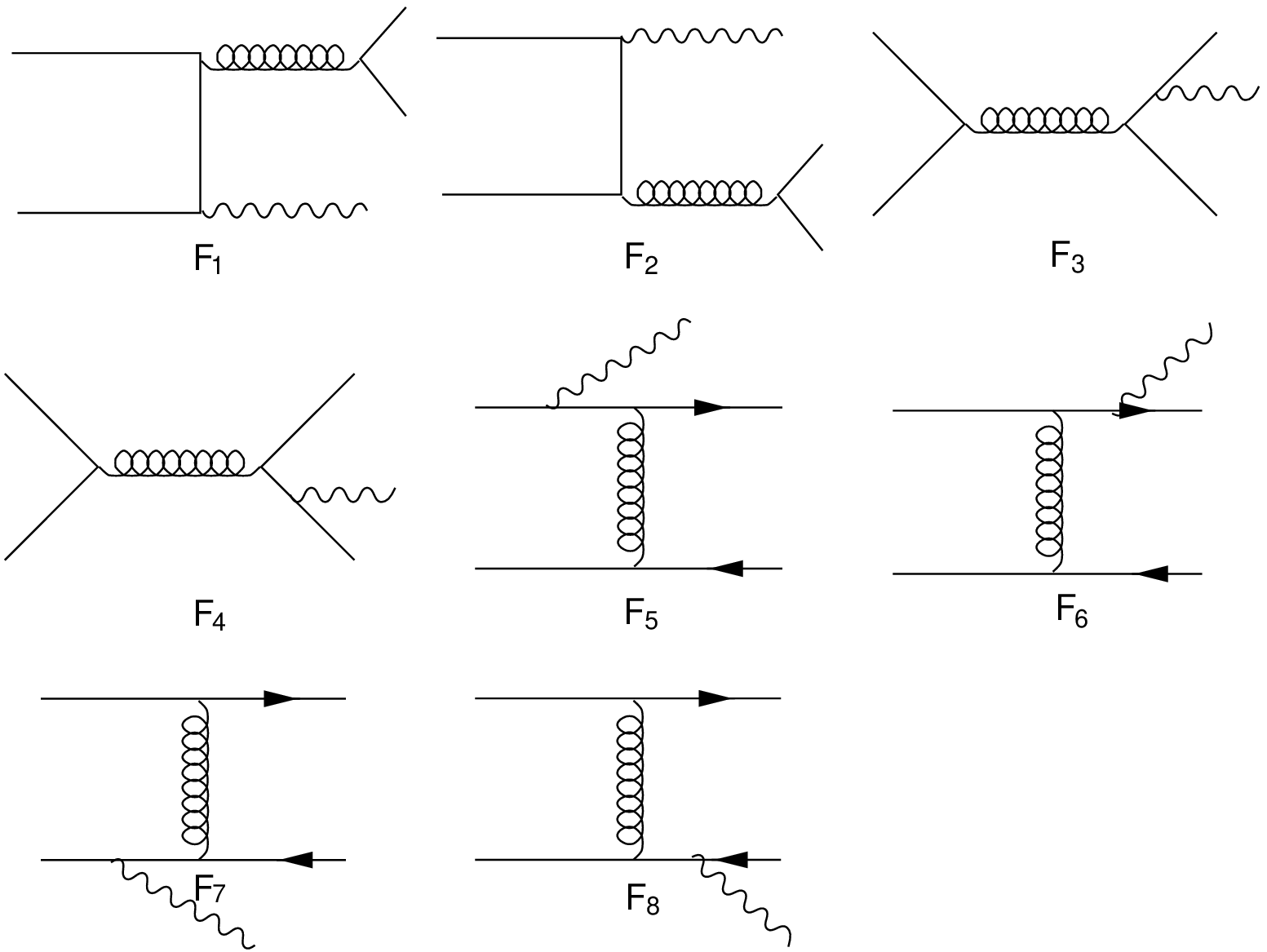}}
\caption{Diagrams which contribute to the process
$q+\bar{q}\rightarrow \gamma^* +q+\bar{q}$}
\label{ffigs}
\end{figure}

The distribution cross section $d\sigma/dt\, du$
starts at order $\alpha_s$ and the two diagrams $L_1$ and $L_2$ denote its 
two Born level contributions in the non-singlet sector, and proceeds to NLO 
(or $O(\alpha_s^2)$) through the radiative corrections of 
Figs.~\ref{ffigs}--\ref{vfigs}. 

The virtual and the real corrections to this cross sections are the same corrections 
that appear in the NNLO analysis of $\sigma( Q^2)$, the invariant mass distribution. 

The extension of $d\sigma/dt\,du $ to the small $q_T$ region can be performed by a resummation \cite{CSS} in impact parameter. 
 At large $q_T$, instead, the cross section is perfectly well defined. 

If we use Dimensional Regularization (DR) (with $n=4 - 2\epsilon$) 
to calculate both the real and the 
virtual corrections to this cross section, then the overlapping 
soft and collinear regions of integration will produce double poles 
$1/\epsilon^2$ and the collinear regions single $1/\epsilon$ poles. 
The study of this cross sections in the non-singlet sector shows that 
``factorization'' takes place both for the unpolarized \cite{EMP} and for the 
longitudinally polarized case $\cite{CCFG}$. By ``factorization'' 
here we mean that the double poles will cancel between the real and the virtual contributions, while the left over collinear singularities ``factorize'', 
i.e. can be absorbed into universal 
parton distributions, which are process independent.  
The usual factorization picture involves also a collinear expansion of the momenta of the hard scatterings around a light-cone direction, fixed by the large 
momenta of the two colliding hadrons, and arrested at a certain order. 
Parton distributions of various twist, in this picture, are defined as matrix elements of nonlocal 
operators at light-like separations. Natural ways to incorporate into 
this picture contributions of higher twist and 
proofs of factorization \cite{QS} have also been formulated. 

An important part of the analysis is the proof of the gauge invariance of the factorization formula up to a certain twist. 
The operatorial expansion of \cite{JJ} shares 
the same features of ref.~\cite{QS}, although the analysis is 
performed directly on the light-cone, rather than in momentum space \cite{QS}. 
 
In both formulations, the violations to the scaling behaviour of the 
twist 2 (and of any twist) structure functions are purely logarithmic, at any order in perturbation theory and all the hard scatterings -for distributions 
of any twist-  are calculated in the 
collinear approximation. 

These approaches, which are manifestly gauge invariant, 
do not isolate any transverse (explicit) $q_T$ in the dominant cut-diagram 
describing the factorization formula. 
Isolation of a fixed $q_T$ in the parton distributions brings the operatorial expressions that describe these distributions away from the light cone.  

In the following, we will assume that 
the leading cut-diagram describing the Drell Yan process at nonzero $p_T$ contains a hard scattering of the form $L_1$ or $L_2$. Therefore, in the leading twist approximation, the $q\bar{q}$ partons undergo hard scattering 
with purely collinear light-like momenta. The $q_T$ of the photon, therefore, is not ``intrinsic'', as 
discussed in \cite{RS,TM}, but it is due to the real (lowest order) emission of a gluon. 
The scaling ($\ln q_T$) violations to this basic picture are generated by the additional 
radiative corrections of Figs.~4-7.   

\section{ $\sigma( Q^2)$ versus $d\sigma/dt\,du$}

In order to obtain $\sigma( Q^2) $ from $d\sigma/dt\,\, du$, then we need to integrate the latter cross section in $t$ and $u$. By doing so, we will generate 
additional singularities (up to $1/\epsilon^4$) after integration, 
since the transverse momentum distribution contains matrix elements of the form $1/t$, $1/u$,  $1/u\,\,t$ etc, which will be responsible for higher order poles 
in $\epsilon$. These higher order singularities, 
according to factorization, will cancel in the final sum of real and virtual, 
after adding the contributions coming from the two-loop on-shell quark form 
factor. 

Therefore, the 
$O(\epsilon^0)$ radiative corrections to the cross section $d\sigma/dt\,\, du$ 
gives, after $t-u $ integration, only the $bulk$ of the total NNLO 
corrections to the invariant mass distribution. These corrections, 
in the longitudinally polarized case, have been calculated \cite{CC}.

\section{ $d\sigma/d^4q\,\, d\Omega$ and transverse spin at nonzero $P_T$}

In their pioneering paper on the polarized Drell Yan process, 
Ralston and Soper 
\cite{RS} introduced the notion of transverse parton distributions. 
It is by now well established that $h_1(x)$ is a leading twist 
(or twist 2) distributions which can be measured in p-p collisions 
using transversely polarized beams. 

As we are going to see, in order to have a non vanishing projection on the 
transverse spin distribution of the hadronic tensor, the parton model prediction requires that the lepton pair should not be completely integrated out.   
The result is common to all the approaches \cite{RS,JJ,TM} followed so far in the analysis of the parton distributions in polarized collisions.

Here we briefly overview these results in order to make our treatment self contained.

The factorized (parton model) picture of the collision emerges 
in an infinite momentum frame, with a large boost parameter 
(formally $P\to \infty$). 

In this limit the two colliding hadrons have large ``plus'' or ``minus'' components and fields are effectively quantized at equal light-cone time $x^+$. 
The degrees of freedom of a fermion are halved, with the negative 
$\psi_-=1/2\gamma^-\gamma^+$ (bad or dependent)
light-cone projections of the Dirac fermions functions of the positive 
$\psi_+=1/2\gamma^+\gamma^-$ (good or independent) ones \cite{JJ,Jaffe1}, 
and of the transverse components of the gauge field.
The plus components can be labeled by helicity eigenvalues 
$\psi_+^{\pm 1/2}$ and chirality eigenvalues of the same sign.   
The negative projections are basically interpreted as quark-gluon composites. 
The requirement that the helicity of the composite be $\pm 1/2$ forces 
the composite to be of the form $\psi_-^{1/2}\sim \psi_+^{-1/2}A^{+1}$
and $\psi_-^{-1/2}\sim \psi_+^{+1/2}A^{-1}$. The chirality of the 
bad quantum components have flipped sign compared to the 
helicities. While the distributions $f_1(x)$ and $g_1(x)$ have a natural interpretation (i.e. are diagonal) in the chirality/helicity basis, the distribution 
$h_1(x)$ does not. Since a fermion of fixed transverse spin is in an eigenstate of the Pauli-Lubanski operator, it is convenient to expand the components of the fermion in a basis of eigenstates of such operator (transversity basis), 
which are expressible as linear combinations of the helicities.  $h_1(x)$ then measures the number of quarks polarized in the transverse direction minus the number of those polarized in the opposite direction in a transverse 
polarized hadron (at a fixed light-cone momentum fraction x). 

One starts by defining the hadronic tensor 

\beq
W^{\mu\nu}=\int {d^4 x\over 2 (\pi)^4}e^{i q\cdot x} 
\langle P_A S_A;P_B S_B|\left[J^\mu(0),J^{\nu}(x)\right]| P_A S_A;P_B S_B
\rangle,
\eeq

with $P_A$ and $P_B$ being the momenta of the colliding hadrons and 
$S_A$ and $S_B$ their covariant spins. 
In a collinear basis 
\beqa
&& P_A= P_A^+ n^+ + {M^2\over 2 P_A^+} n^-,\nonumber \\
&& P_B= P_B^- n^- + {M^2\over 2 P_B^-} n^+, \nonumber \\
&& S_A= {\lambda_A P_A^+\over M}n^+ - {\lambda_A M\over 2 P_A^+}n^-,\nonumber \\
&& S_A= {\lambda_B P_B^-\over M}n^- - {\lambda_B M\over 2 P_B^-}n^+,
\eeqa
with spin vectors orthogonal to the hadron momenta. 

$P_A^+$ ands $P_B^-$ are the two large light-cone components of the incoming 
hadrons, of equal mass $M$. 

Assuming factorization, the quantum numbers of the two hadrons 
can be decoupled by Fierz transformations, and 
the interaction described at leading order by a single hard scattering, with a single photon in the unitarity diagram.  

The distribution functions that emerge -at leading order- 
from this factorized picture are 
correlation functions of non-local operators in configuration space. 
They are the quark-quark and the quark-antiquark 
correlators.

Their expression simplifies in the axial gauge, in which the 
gauge link is removed by the gauge condition. For instance, the quark-quark correlator takes 
the form 

\beq
\left(\Phi_{a/A}\right)_{\alpha\beta}(P,S,k)=\int {d^4z\over (2 \pi)^4} 
e^{i k\cdot z}
\langle P,S|\overline{\psi}^{(a)}_\beta (0)\psi^{(a)}_\alpha(z)|PS\rangle.
\label{bu}
\eeq
We have included the quark flavour index $a$ and an index $A$ for the hadron, 
as usual. 
Fields are not time ordered since they can be described by the good 
light cone components $\psi_+$ and $A_T$, as discussed in \cite{RLJ}. 
 
Using hermicity, parity and time reversal invariance, one can give a 
general expansion for these correlators in terms of basic amplitudes which are 
functions of P,S, k and are Dirac algebra valued.  

However, the basic matrix elements that appear in leading power factorization are not the correlators themselves, but their integrals over the $k^-$ and $k_T$ components of momentum. 

Specifically, one introduces the Sudakov expansion
\beq
 k=x p^+ n^+ + {k^2- k_T^2\over 2 x}n^- +k_T 
\eeq

and may decide to integrate over both $k^-$ and $k_T$ in (\ref{bu}).

Using the identity 
\beq
\int {d(k\cdot p) d^2k_T\over (2 \pi)^4}e
^{i k\cdot z} =\int {d\lambda\over 2 \pi} \delta^4(z - \lambda n) e^{i \lambda x},
\eeq
and projecting over the Dirac basis of the 16 independent $\gamma$ matrices, 
one obtains the general expansion of the correlator
\beqa
&& \int\phi= {\bf 1} {1\over 4}\int Tr\left[\Phi\right]
- i \gamma_5\int Tr\left[i \gamma_5 \Phi\right] + \gamma_\mu 
 {1\over 4}\int Tr\left[\gamma^\mu \Phi\right]\nonumber \\
&&\hspace{2 cm}  +\gamma_5\gamma_{\mu}{1\over 4}
Tr\left[\gamma_\mu\gamma_5\Phi\right] +i\gamma_5\sigma^{\mu\nu}\int 
Tr\left[i \sigma^{\mu\nu}\gamma_5 \Phi\right], 
\eeqa
where the sign of integration $\int $ is a shorthand notation for the $k^-, k_T$ 
integration in $k$

\beqa
 \int Tr\left[\Gamma \Phi\right] &\equiv& 
{1\over 2}\int d(k\cdot p)d^2 k_T Tr\left[\Gamma\Phi(k)\right] \nonumber \\
&=&\int {d\lambda\over {4\pi}} e^{i\lambda x }
\langle P,S|\overline{\psi}(0)\Gamma \psi(\lambda n)|PS\rangle.
\eeqa

Therefore, after integration, the correlators are sampled on light-cone 
rays. 

The transverse spin distribution $h_1(x)$ appears in the projector of the 
quark-quark correlator over $i \sigma^{\mu\nu}\gamma_5$

\beqa
\int {d\lambda\over 4 \pi}e^{i\lambda x}\langle PS|
\overline{\psi}(0) i \sigma^{\mu\nu}\gamma_5\psi(\lambda n)|P S\rangle 
&=&(S_T^\mu p^\nu - S_T^\nu p^\mu) h_1(x)\nonumber \\  
&&\hspace{-3.5cm}+ M\lambda(p^\mu n^\nu - p^\nu n^\mu)h_L(x)
 + M^2(S_T n^\nu - S_T^\nu n^\mu)h_3(x).
\eeqa

Using Jaffe's definition of twist \cite{Jaffe1} $t=2 +$number of $M$, one finds that 
$h_1(x)$ is twist$-2$, $h_L(x)$ is twist$-3$ ands $h_3(x)$ is twist$-4$. 

The leading twist expansion of the quark-quark correlator then is of the form

\beq
\int {d\lambda\over 2 \pi}e^{i\lambda x}\langle PS|
\overline{\psi}(0)\psi(\lambda n)|P S\rangle={1\over 2}
\left( \not{p}f_1(x) +\lambda \gamma_5 \not{p} g_1(x) +
\gamma_5 \not{S_T}\not{p}h_1(x)\right).
\eeq

In their parton model analysis the authors of \cite{JJ} obtain expressions for the 
hadron tensor $W^{\mu\nu}$ in terms of parton distributions which are 
peaked at $q_T=0$, (eqs (67), (68) and (69) of 
\cite{JJ}). 

A more general approach has been discussed in \cite{TM}. 
There, the authors, following Ralston and Soper, assume a factorization 
with an intrinsic $q_T$ for the total momentum of the partons entering hard scattering. 
The factorized expression for the hadronic tensor is assumed of the form 
\cite{RS,TM}
\beqa
 W^{\mu\nu}&=&{1\over 2}\sum_{a b}\delta_{b \bar{a}}e_a^2\int d^4 k_a d^4 k_b 
\delta^4(k_a + k_b- q)\nonumber \\
&& \,\,\,\,\,\,\,\,\times Tr[\Phi_{a/A}(P_A S_A,k_a)\gamma^{\mu}
\overline{\Phi}_{b/B}(P_B,S_B,k_b)\gamma^\mu].
\label{ff}
\eeqa
Notice that this primordial $q_T$ is not 
induced by radiative corrections. In this last approach, by setting $q_=0$ 
(or $q_T=0$) one is yet unable to recover the results of Jaffe and Ji, since the 
integrals over the parton momenta $k_A$ and $k_B$ in the hard scattering are still coupled. They can be uncoupled only if one starts with a general $q_T$ 
in the factorization formula Eq. (\ref{ff}), and integrates over $q_T$. 
Therefore, the approaches of \cite{TM,JJ} 
compare at the level of the integrated cross section $d\sigma/dq^+ dq^- d\Omega$, an expression derived originally in 
\cite{RS}. The authors of \cite{TM} also claim 
to be in disagreement with the expression quoted in \cite{JJ} for the $A_{LT}$ or 
longitudinal-transverse asymmetry. 

The parton level results for the $q_T$-integrated cross sections, 
however, in both approaches, agree, and are also in agreement with the result given by Ralston and Soper, that we simply quote

\beqa
{d\sigma\over  dq^+ dq^- d\Omega}&=&
{ \alpha^2\over 4 Q^2}{1\over 3}\sum e_Q^2 f_1(x_A)f_1(x_B) \nonumber \\
& \times & 
(1- \cos^2\theta)(1- \lambda_A \lambda_B g_1(x_A) g_1(x_B))\nonumber \\
&&+ \sin^2\theta \cos(2 \phi - \phi_A -\phi_B)|S_A^T| |S_B^T| 
h_1(x_A) h_1(x_B).
\label{bebe}
\eeqa

The angles $\theta$ and $\phi$ in Eq. (\ref{bebe}) 
refer to the leptons and are measured respect 
to the Collins-Soper (see below) frame, while $\phi_A$ and $\phi_B$  are the azimuthal angles of the two transverse spins 4-vectors in the same frame. 

It clearly shows that an integration over the azimuthal angle of the lepton pair makes the projection over $h_1(x)$ vanish.

\section{Parton Level Analysis}

Since the structure of the final state in the analysis of the transverse spin  
cannot be neglected, it is convenient to set up some diagrammatic notations 
and divide all the calculations into two subsets. Diagrammatically, this is 
easily understood in terms of cut graphs. 
For instance, the contraction of the hadronic tensor with the diagonal 
part (i.e., with $g_{\mu\nu}$) of the leptonic tensor 

\beq
L_{\mu\nu}\sim v_{1\mu}v_{2\nu} + v_{2\mu}v_{1\nu} - g_{\mu\nu}v_1\cdot v_2
\eeq
is described in terms of cut diagrams of 
cross sections for the  process  $q\bar{q}\to \gamma^*$. 

The non diagonal part, instead, is schematically described by cut diagrams 
of cross sections for the process $q\bar{q}\to l^+ l^-$.

In the next sections we will analyze the NLO structure of the non-diagonal non-singlet virtual corrections for the $p_T$ distributions.

Notice that if we integrate completely over the final leptons and use the 
gauge invariance of the partonic matrix elements, 
then only the diagonal projection survives.

In our calculation we use Dimensional Regularization 
(DR) all the way. The chiral projectors are regulated 
according to the t'Hooft-Veltman prescription (HV) \cite{HV}.
  
There are clearly many advantages in being able to use this regularization, 
given the complexity of the calculations. 
However this is not generally possible in most of cases, except for 
specifically defined cross sections. As we have just mentioned, the cross section has to be sufficiently inclusive in order to be able to cancel 
the infrared divergences order by order in perturbation theory. 
At the same time, in the case of transverse spin, 
we need to leave in part the lepton pair in the final state unintegrated.
As we are going to show in the next sections, one possible choice is given by the cross section $d\sigma/d^4q \,\,d\Omega$, with $\Omega$ evaluated in the 
rest frame of the photon, which we are going to investigate.

\section{ About the treatment of $\gamma_5$}
Here are some remarks concerning the extension used for $\gamma_5$ 
in n dimensions.
In the t'Hooft-Veltman prescription for $\gamma_5$ for $n=4 -2 \epsilon$ dimensions, the 
Dirac algebra is split into a 4 and into an $n-4$ subspace. 

The metric space is also split in a similar way

\beq
g^{\mu\nu}=\widehat{\widehat{g}}_{\mu\nu} +\widehat{g}_{\mu\nu},
\eeq

and the Dirac algebra written as a direct sum 

\beq
\gamma_\mu=\widehat{\widehat{\gamma}}_\mu + \widehat{\gamma}_\mu. 
\eeq

The anticommutation relations for the $\gamma$ matrices are

\beq
 \left[\widehat{\widehat{\gamma}}_\mu,\widehat{\widehat{\gamma}}_\nu\right]_+=2 \widehat{\widehat{g}}^{\mu\nu},\hspace{.5cm}
\left[{\widehat{\gamma}}_\mu,{\widehat{\gamma}}_\nu\right]_+=2 
{\widehat{g}}^{\mu\nu},\hspace{.5cm}
\left[{\widehat{\gamma}}_\mu,\widehat{{\widehat{\gamma}}}_\nu\right]_+=0.
\nonumber \\
\eeq
The antisymmetric tensor $\epsilon_{\mu\nu\rho\sigma}$ is 4 dimensional and 
projects only over the 4 dimensional subspace

\beq
\epsilon_{\mu\nu\rho\sigma}\widehat{g}_{\sigma\eta}=0,\hspace{1.5cm}
\epsilon_{\mu\nu\rho\sigma}\widehat{\gamma}_{\sigma}=0.
\eeq

$\gamma_5$ is defined  in n-dimensions as 

\beq
\gamma_5=
{i\over 4!} \epsilon^{\mu\nu\rho\sigma}\gamma^\mu\gamma^\nu\gamma^{\rho}\gamma^{\sigma}.
\eeq

With this definition, $\gamma_5$ anticommutes with the 4-dimensional 
$\gamma$'s and commutes with the remaining the n-4-dimensional ones.

\section {Distributions at nonzero $p_T$}

We calculate the spin dependence of the cross section using the 
relation \cite{collins}  

\beq
u(p,s)\bar{u}(p,s)={1\over 2}\slash{p}(1-\lambda\gamma_5+\gamma_5
\slasher{s}_{\perp}) \ ,
\eeq
where $s_{\perp}$ represents the transverse components of the spin vector 
$s^\mu$,
and $s^{\mu}s_{\mu}=-1$.  Clearly, $\lambda$ and $|s_{\perp}|$
are not independent, and in fact satisfy the relation $\lambda^2+ 
|s_{\perp}|^2 = 1$.

%We also set $h_1\,=\,\pm 1\,=2 \,\lambda_1$ and $h_2\,=\,\pm 1\,
% 2 \lambda_2$, with $\lambda_1$ and $\lambda_2$ being the 
%two helicities of the quark and of the antiquark respectively. We also set 
%$h\,=\,h_1\,h_2$.

In our NLO calculation $\alpha_s\to\alpha_s(\mu^2)$ is the
running coupling constant renormalized at the scale $\mu^2$ in the
$\overline{MS}$ scheme, satisfying the renormalization group equation
\begin{equation}
\mu^2\frac{d\alpha_s}{d\mu^2}=-\alpha_s\left[
\beta_0\frac{\alpha_s}{4\pi}+\beta_1\left(
\frac{\alpha_s}{4\pi}\right)^2+O(\alpha_s^3)\right] \ ,
\end{equation}
with
\begin{eqnarray}
\beta_0&=&\frac{11 N_C}{3}-\frac{2 N_F}{3} \nonumber \\
\beta_1&=&\frac{34 N_C^2}{3}-\frac{10 N_C N_F}{3}-2 C_F N_F \ ,
\end{eqnarray}
and with $N_F$ the number of active quark flavors.

The lowest order ({\it non-singlet}) contributions to the real emissions 
arise from the two, $q+\bar{q}\to 
\gamma^*+g$,  amplitudes $L_1$ and $L_2$.
As mentioned before, they are also the lowest order 
contributions to the $p_T$ distributions in Drell Yan. 

We refer 
to these two diagrams as the {\it direct} and the {\it exchange} 
(or crossed) amplitudes, respectively. 

The squares of the direct amplitude $L_1$ and exchange amplitude 
$L_2$ in Fig.~\ref{low} in $n=4-2\epsilon $ dimensions are given by
\beqa
M_{dd}(s_1,s_2)&=&e_f^2g^2g_s^2 {C_F\over N_c}
{2u\over t}\left[ (1-\eps)^2-h(1+\eps)^2 \phantom{\frac{1}{1}} \right.\nonumber\\
&&\left.\hspace{1cm}-\eps^2\left(s_{1 \perp}\cdot s_{2 \perp} + {2s \over tu}
(k_2 \cdot s_{1 \perp})(k_2 \cdot s_{2 \perp})\right)\right],\nonumber \\
M_{cc}(s_1,s_2)&=&e_f^2g^2g_s^2 {C_F\over N_c}
{2t\over u}\left[ (1-\eps)^2-h(1+\eps)^2 \phantom{\frac{1}{1}} \right.\nonumber\\
&&\left.\hspace{1cm}-\eps^2\left(s_{1 \perp}\cdot 
s_{2 \perp} + {2s \over tu} (k_2 \cdot s_{1 \perp})(k_2 
\cdot s_{2 \perp})\right)\right].
\eeqa
$e_f$ the 
charge of the quark, and we have set $\lambda_1 \lambda_2 = h$.  
The quantity $C_F/N_c$ is the 
color factor, and $\alpha_s=g_s^2/4\pi$ is the QCD strong 
coupling constant. 

The interference term is more complicated 
and is given by,
\beqn
&& 2M_{dc}(s_1,s_2)= \nonumber \\
&& e_f^2g^2g_s^2
{C_F\over N_C}
{4\over tu}\left[ (1-\eps)(Q^2s-\eps tu)-h(1+\eps)(Q^2s+\eps tu)
-2h\eps tu \right.\nonumber \\
&&\left.+(s_{1 \perp}\cdot s_{2 \perp})((Q^2s+tu)-(1+\eps)(Q^2s+\eps tu))
\right.\nonumber \\
&&\left.+\left({(k_2 \cdot s_{1 \perp})(k_2 \cdot s_{2 \perp}) \over Q^2}
\right)
(2sQ^2(1-\eps-\eps^2))\right].
\eeqn
The sum of the two Born amplitudes squared is 
\beqn
&& |M_{B}(s_1,s_2)|^2=M_{dd}(s_1,s_2)+2M_{dc}(s_1,
s_2)+M_{cc}(s_1,s_2)\nonumber \\
&& =e_f^2g^2g_s^2
{C_F\over N_C}
{2\over tu}\left[(1-h)(1-\eps)\left(2Q^2s+(1-\eps)(t^2+u^2)-2\eps
tu\right)\right.\nonumber \\
&&\,+4h\eps \left(Q^2s+(u+t)^2\right)\nonumber \\
&&\,+(s_{1 \perp} \cdot s_{2 \perp})\left(2tu-2\eps(s+u)(s+t))-
\eps^2(t+u)^2\right)\nonumber \\
&&\left.+\left({(k_2 \cdot s_{1 \perp})(k_2 \cdot s_{2 \perp})
\over t u}\right)
2s(2tu(1-\eps)-\eps^2(t+u)^2)\right].
\label{bbo}
\eeqn

The distribution cross section is defined by 
\begin{equation}
s{d\hat\sigma_B\over dt}=e_f^2K_2{\alpha_s\over
s}\left[T_1(Q^2,u,t,h)+T_2(Q^2,u,t,s_{\perp1},s_{\perp2})\right].
\label{bornddt}
\end{equation}
$K_2$ is defined by
\beq
K_2=\pi\alpha{C_F\over N_C}
{1\over\Gamma(1-\eps)}
\left({4\pi\mu^2\over Q^2}\right)^\eps\left({sQ^2\over
tu}\right)^\eps,
\eeq
where we have rescaled, $\alpha_s\to\alpha_s(\mu^2)^\eps$, so
that it remains dimensionless in $n=4-2\eps$ dimensions.
The first component of the cross section is the same as longitudinally
polarized cross section
\begin{equation}
T_1(Q^2,u,t,h)= (1-h) T_B(Q^2,u,t) + 8h\epsilon \frac{Q^2 s +(u+t)^2}{t u}
\end{equation}
where $T_B$ is the unpolarized born level matrix element.
\begin{equation}
T_B(Q^2,u,t)= 2 (1-\epsilon)\left[
(1-\epsilon)\left(\frac{u}{t}+\frac{t}{u}\right) +
\frac{2Q^2(Q^2-u-t)}{ut}- 2 \epsilon\right].
\end{equation}
The transverse spin contribution to the cross section is
\begin{eqnarray}
T_2(Q^2,u,t,s_{1\perp},s_{2\perp})&=& 2 s_{1\perp}\cdot s_{2\perp}
\frac{2 t u - 2 \epsilon (s+u) (s+t)  - \epsilon^2 (t + u)^2}{t u}
\nonumber\\
&&+ 4 (k_1\cdot{s}_{1\perp}) (k_1\cdot{s}_{2\perp})
\frac{s (2(1  - \epsilon) t u - \epsilon^2 (t + u)^2)}{t^2 u^2}
\nonumber\\
&=&
2\beta \left(2 (1 -  \epsilon)  - \epsilon^2 \frac{(t +
u)^2}{t u}\right)
- 4\epsilon  (s_{1\perp}\cdot s_{2\perp}) \frac{s Q^2}{t u},
\end{eqnarray}
where
\begin{equation}
\beta= \frac{2(k_1\cdot{s}_{1\perp}) ({k}_1\cdot{s}_{2\perp}) s}{t u}
+{s}_{1\perp}\cdot{s}_{2\perp} .
\end{equation}
If $s_{1\perp}=s_{2\perp}=0$, Eq.~(\ref{bornddt}) reduced to the 
longitudinally polarized cross section cross section of Ref.~\cite{CCFG}.

In the limit $h =0 $
\begin{equation}
\vec{s}_{1\perp}\cdot\vec{s}_{2\perp}=- s_{1\perp}\cdot s_{2\perp}
 = \cos\Delta\theta
\end{equation}
and
\begin{eqnarray}
\frac{2(k_1\cdot{s}_{1\perp}) ({k}_1\cdot{s}_{2\perp}) s}{t u}
+{s}_{1\perp}\cdot{s}_{2\perp} 
&=& \frac{2(\vec{k}_1\cdot\vec{s}_{1\perp}) (\vec{k}_1\cdot\vec{s}_{2\perp}) s}{t u}
-\vec{s}_{1\perp}\cdot\vec{s}_{2\perp} 
\nonumber\\
&=& 
\frac{2 k_\perp^2 s}{tu} \cos\theta_1 \cos(\theta_1+\Delta\theta) 
-\cos\Delta\theta
\nonumber\\
&=& 
\cos(2\theta_1+\Delta\theta).
\end{eqnarray}
Here,  $k_\perp^2= (t u)/s $ and $\theta_1$ is an angle between
$\vec{k}_\perp$ and $\vec{s}_{1\perp}$. 

Away from 4 dimensions $T_1$ clearly contains helicity 
violating contributions, due to the prescription chosen for $\gamma_5$.

\section{The Structure of the Non Singlet}

The radiative corrections to the non singlet come from the interference between
the two diagrams $L_1$ and $L_2$ and the diagrams denoted $V_1,....V_{11}$

\beq
d{\hat\sigma}^{virtual}\sim 
2Re\left((L_1+L_2)\sum_{i=1}^{11}V_i^*\right),
\label{rdf_ampv}
\eeq

The other real $\bar q_fq_f$ diagrams are
\beq
d{\hat\sigma}^{real}_{\bar q_fq_f}\sim
|\sum_{i=1}^8F_i|^2+|\sum_{i=1}^8G_i|^2,
\eeq
where the amplitudes $F_i$ are shown in Fig.~4 and the amplitudes $G_i$ are shown in Fig.~5 with $f=f'$.  The absolute square of the $F_i$ amplitudes 
can be written as:
\begin{eqnarray}
\lefteqn{|\sum_{i=1}^8F_i|^2=}\nonumber \\
&&|F_1+F_2|^2+|F_3+F_4|^2+2Re\left(\sum_{i=1}^4F_i\sum_{i=5}^8F^*_i\right)
+|\sum_{i=5}^8F_i|^2,
\end{eqnarray}
having used
\beq
Re\left((F_1+F_2)(F_3+F_4)^*\right)=0.
\eeq
Therefore, the diagonal real $\bar q_fq_f$ contributions are
\beq
d{\hat\sigma}^{real}_{\bar q_fq_f}= d{\hat\sigma}_1
+d{\hat\sigma}_2+d{\hat\sigma}_3+d{\hat\sigma}_F,
\label{roro1}
\eeq
where
\begin{eqnarray}
d{\hat\sigma}_1&\sim& |F_1+F_2|^2+|\sum_{i=1}^8G_i|^2,\label{rdf_amp1}\\
d{\hat\sigma}_2&\sim& |F_3+F_4|^2,\label{rdf_amp2}\\
d{\hat\sigma}_3&\sim& 2Re\left(\sum_{i=1}^4F_i\sum_{i=5}^8F^*_i\right),\label{rdf_amp3} \\
d{\hat\sigma}_F&\sim& |\sum_{i=5}^8F_i|^2.
\end{eqnarray}

The real diagonal $q_f\,q_f$ diagrams are
\beq
d{\hat\sigma}^{real}_{q_fq_f}\sim
|\sum_{i=1}^4H_i-\sum_{i=4}^8H_i|^2,
\label{roro2}
\eeq
where the amplitudes $H_i$ are shown in Fig.~6 and the relative minus sign between the direct and exchange diagrams is due to Fermi statistics since $f=f'$. The absolute square of the $H_i$ amplitudes can be written as follows:
\begin{eqnarray}
\lefteqn{|\sum_{i=1}^4H_i-\sum_{i=4}^8H_i|^2=} \nonumber \\
&&|\sum_{i=1}^4H_i|^2+|\sum_{i=4}^8H_i|^2
-2Re\left(\sum_{i=1}^4H_i\sum_{i=4}^8H_i^*\right).
\end{eqnarray}
Thus the diagonal real $q_fq_f$ contributions are,
\beq
d{\hat\sigma}^{real}_{q_fq_f}= 
d{\hat\sigma}_H+d{\hat\sigma}_4,
\label{rdf_nsqq}
\eeq
where
\begin{eqnarray}
d{\hat\sigma}_H&\sim& |\sum_{i=1}^4H_i|^2+|\sum_{i=4}^8H_i|^2,\\
d{\hat\sigma}_4&\sim& -2Re\left(\sum_{i=1}^4H_i\sum_{i=4}^8H_i^*\right)\label{rdf_amp4}.
\end{eqnarray}
The diagonal ($f=f'$) real part of the non-singlet cross section 
is given by subtracting
(\ref{roro2}) from (\ref{roro1}) as follows:
\beq
d{\hat\sigma}^{real}_{NS}=d{\hat\sigma}_1
+d{\hat\sigma}_2+d{\hat\sigma}_3-d{\hat\sigma}_4,
\label{rdf_ns3}
\eeq
where we have used the fact that $d{\hat\sigma}_F=d{\hat\sigma}_H$ which arises due to
\beq
\int_{PS_3} |\sum_{i=5}^8F_i|^2
=\nicefrac12 \int_{PS_3} |\sum_{i=1}^4H_i|^2
+\nicefrac12 \int_{PS_3}|\sum_{i=4}^8H_i|^2.
\eeq
When integrating over the two identical particles $k_3$ and $k_4$ over the phase space $PS_3$ in (\ref{ps1}) an extra statistical factor of $1/2$ must be inserted so as not to double count.  Thus, the diagonal ($f=f'$) part of the 
complete non-singlet cross section is given by
\beq
d{\hat\sigma}_{NS}=d{\tilde\sigma}_1
+d{\hat\sigma}_2+d{\hat\sigma}_3-d{\hat\sigma}_4,
\label{rdf_ns4}
\eeq
where
\beq
d{\tilde\sigma}_1=d{\hat\sigma}^{fact}_1+d{\hat\sigma}^{virtual},
\eeq
and where $d{\hat\sigma}^{fact}_1$ is the ``factorized" cross section given by
\beq
d{\hat\sigma}^{fact}_1=d{\hat\sigma}_1+d{\hat\sigma}^{counter}_1.
\label{rdf_counter}
\eeq
It is necessary to remove the collinear singularities from the initial state 
by subtracting the cross section
$d{\hat\sigma}^{counter}_1$ to $d{\hat\sigma}_1$.

The off-diagonal real $\bar q_fq_{f'}$ diagrams are
\beq
d{\hat\sigma}^{real}_{\bar q_fq_{f'}}\sim
|\sum_{i=1}^4F_i|^2,
\eeq
where the amplitudes $F_i$ are shown in Fig.~4 and where $f\neq f'$.
Similarly, the off-diagonal real $q_fq_{f'}$ diagrams are
\beq
d{\hat\sigma}^{real}_{q_fq_{f'}}\sim
|\sum_{i=1}^4H_i|^2,
\eeq
where the amplitudes $H_i$ are shown in Fig.~6.  In this case,
\beq
|\sum_{i=1}^4F_i|^2=|\sum_{i=1}^4H_i|^2,
\eeq
and these two contributions cancel when forming the non-singlet cross section.

The total contribution to the transversely polarized cross section 
can then be written in the form 

\beq
d\sigma=d\sigma_{NS} + 2\left( d\sigma_H + d\sigma_4\right)_{ff}+
 \left( d\sigma^{real}_{q_f q_{f'}} + 
d\sigma^{real}_{q_f \bar{q}_{f'}}\right),
\eeq

where the suffix $ff$ refers the diagonal (in flavour) 
contribution  and the remaining  term in bracket are the off diagonal 
scattering diagrams in the F and H sets.

 \section{ Off-shell renormalization and tensor reductions }
Although in the non-singlet sector we do not encounter any anomalous diagram, 
the regularization scheme still suffers from unphysical features, such as 
helicity violation, which are absent in other schemes. We should also mention that in the case of the 2-to-2 contributions, all the dependence on the 
hat-momenta of the the hard scattering can be eliminated. This is 
equivalent to ask that the momenta which are not integrated -and also the 
spin 4-vectors- have only 4 dimensional components.
 
The calculation of the radiative corrections to the 
vertices is often performed by using various tricks in order 
to simplify the loop momenta, such as the Feynman parameterization combined with symmetric integration. An alternative way to proceed is to use the 
Passarino-Veltman (PV) \cite{pave} recursion procedure in order to relate the tensor integrals 
which appear in the calculation to the corresponding scalar ones.  
The PV method introduces scaless integrals which have to be handled with particular care. In an appendix we illustrate in 
a simple way how to obtain the renormalized expression of all the coefficients of the expansion of tensor integrals to scalar form. The approach \cite{CC} is
 easy to 
implement in symbolic manipulations. We start from the expression of any tensor integral 
and apply the Passarino Veltman (PV) reduction procedure to determine the coefficients of the expansion. Only the renormalization of the scalar self energy 
diagram $B_0(Q^2)$ (and of its $Q^2\to 0$ limit) is needed. We use dimensional regularization with a single 
parameter $\epsilon$ to regulate both UV and IR singularities. 
Notice that scaless integrals such as $B_0(0)$, so generated, 
are intrinsically ambiguous. 
While it is possible to set such terms to zero by definition  
(on-shell), off-shell renormalization of $B(Q^2)$ leaves us with a $1/\epsilon$ pole contribution in 
$B_0(0)$ after that the $Q^2\to 0$ limit is performed. We isolate and 
remove the UV singularity (by setting $n=4 - 2 \epsilon$), we then switch 
$\epsilon\to -\epsilon$ to regulate the remaining IR divergences. Finally, we 
send $Q^2\to 0$. This leaves us with a $1/\epsilon$ pole for $B_0(0)$, of IR 
origin. Details of the method can be found in the appendix.

\section{NLO Diagonal Virtual Corrections}

In this section we present results for the diagonal corrections to the NLO 
non-singlet cross section and show their factorization. Although the evaluation of these diagrams is a formidable task, the result can be condensed 
in a remarkably simple form. We get 

\begin{eqnarray}
\!\!&&\!\!\!\!\!\!\!\!\!\frac{s d\sigma^{\rm virtual}}{dtdu} = e^2_f K_2\frac{\alpha_s}{s}\delta(s+t+u-Q^2)
\left\{ \left(T_1+ T_2\right) 
\left[ 1 - \frac{\alpha}{2\pi}\frac{\Gamma(1-\epsilon)}{\Gamma
(1- 2 \epsilon)}\left(\frac{4\pi \mu^2}{Q^2}\right)^{\epsilon}\right.\right.
\nonumber \\
 &  & \left.\ \times\left(\frac{2 C_F + N_C}{\epsilon^2} 
+ \frac{1}{\epsilon}
\left(3 C_F - 2 C_F \ln \frac{s}{Q^2} + \frac{11}{6} N_C + N_C \ln
\frac{sQ^2}{ut} -\frac{1}{3}N_F\right)\right)\right]
\nonumber \\
&&\left.+\frac{\alpha}{2\pi}\left[
(1+h) R_1 + R_2 \right]\right\},
\label{main}
\end{eqnarray}

where
\begin{eqnarray}
\!\!&&\!\!\!\!\!\!\!\!\! R_1 = 
\pi^2(4C_F + N_C)\frac{2Q^2s+t^2+u^2}{3tu}
-2(2C_F- N_C)\frac{Q^2 (t^2 + u^2)}{tu(t +  u)}  
\nonumber \\ &&
 -2 C_F  
\left(\frac{8(2Q^2 s +t^2+u^2)}{tu}  -\frac{Q^4 s(t+u)}{t u
(s+t)(s+u)} -\frac{t^2+u^2}{(s+t)(s+u)}\right)   
\nonumber \\ &&
+\left[\left\{  
  (2 C_F - N_C) Li_2\left(-\frac{t + u}{s}\right)
\frac{2Q^2 s +t^2+u^2+2s^2}{t u} \right.\right.
\nonumber \\ &&
- \left( 2 Li_2\left(\frac{t}{t-Q^2}\right) 
+\ln^2\left(\frac{t-Q^2}{t}\right) \right) 
\left(N_C \frac{2 s+ t}{u} + 2 C_F \frac{s^2+(s+u)^2}{t u}
\right) 
\nonumber \\ &&
+  (2 C_F - N_C)\left(2\ln\left(\frac{s}{Q^2}\right) 
 \frac{Q^4 - (t+u)^2 }{(t + u)^2} 
 + \ln^2\left(\frac{s}{Q^2}\right)\frac{s^2}{t u} 
 \right.
\nonumber \\ &&
\left.
-  \left(2 \ln\left(\frac{|t|}{Q^2}\right)\ln\left(\frac{s}{Q^2}\right) 
- \ln^2\left(\frac{|t|}{Q^2}\right) \right)
     \frac{s^2+(s+u)^2}{t u} 
\right)
\nonumber \\ &&
+ 2 \ln\left(\frac{|t|}{Q^2}\right) \left(C_F \frac{4Q^2 s-2 s t+t u
 }{(s+u)^2 } + N_C  \frac{t}{s+u} 
\right)
\nonumber \\ &&
\left.\left.-N_C\ln\left(\frac{|t|}{Q^2}\right)\ln\left(\frac{|u|}{Q^2}\right) 
\frac{ 2Q^2 s+t^2+u^2}{ t u} \right\} + \{ t \leftrightarrow u\}
\right],
\end{eqnarray}

describe the NLO dependence on the longitudinal components of spin.
The transverse spin contributions are condensed into the expression

\begin{eqnarray}
\!\!&&\!\!\!\!\!\!\!\!\! R_2 = 
 (s_{1\perp}\cdot s_{2\perp})\left[
\frac{11}{3}N_C \frac{Q^2 s}{t u}
+ 2(2C_F - N_C) \left(\ln\left(\frac{|t|}{Q^2}\right) \frac{2 s}{s+u} 
+  \ln\left(\frac{|u|}{Q^2}\right) \frac{2 s}{s+t} \right)\right]
\nonumber \\ &&
+\beta\left[(4C_F + N_C)\frac{\pi^2}{3}
-  2 C_F \left( 9+ \frac{2s}{t+u} \right) - N_C \left(
\frac{11}{3} +\frac{(t+u)^2}{t u}\right)\right.
\nonumber \\ &&
+ \left.\left\{  
  2(2 C_F - N_C)
  Li_2\left(-\frac{t + u}{s}\right)
 \frac{s^2+u^2}{u^2}
\right.\right.
\nonumber \\ &&- \left( 2 Li_2\left(\frac{t}{t-Q^2}\right) 
+\ln^2\left(\frac{t-Q^2}{t}\right) \right) 
\frac{2 C_F (s^2+u^2)- N_C(s^2-u^2)}{u^2}
\nonumber \\ &&
+  (2 C_F - N_C)\left(
  \ln^2\left(\frac{s}{Q^2}\right)
\frac{s^2}{u^2} 
-  \left(2 \ln\left(\frac{|t|}{Q^2}\right)\ln\left(\frac{s}{Q^2}\right) 
- \ln^2\left(\frac{|t|}{Q^2}\right) \right)
      \frac{s^2+u^2}{u^2} 
\right)
\nonumber \\ &&
+ 2 \ln\left(\frac{|t|}{Q^2}\right) \left(
3C_F -(2C_F-N_C)\left( \frac{2 u}{t} + \frac{s+u}{u} +\frac{2 u}{s+u}\right) 
\right)
\nonumber \\ &&
\left.\left.-2 N_C\ln\left(\frac{|t|}{Q^2}\right)\ln\left(\frac{|u|}{Q^2}\right)
\right\} + \{ t \leftrightarrow u\}
\right].
\label{mainv}
\end{eqnarray}

Notice that in the rest frame of the photon, 
$\beta=s_{1\perp}\cdot s_{2\perp}$, and the expression above simplifies 
substantially. From Eq.~ (\ref{main}) it appears obvious that the 
virtual corrections factorize. In fact the double poles $1/\epsilon^2$ correctly multiply the 2-to-2 Born cross section calculated in n dimensions, 
evaluated with the most general spin dependence in the initial state. 
If we 
send any of the transverse spin vectors to zero, 
then the result coincides with the longitudinally polarized corrections obtained in ref.~ \cite{CCFG}. Finally, by sending to zero any of the helicities
and any of the 2 spin vectors, one reobtains the result given in ref.~\cite{EMP} in the unpolarized case. Eq.~(\ref{main}) clearly contains 
(finite) 
helicity violating contributions which are not canceled by the real 
emissions in the $\overline{MS}$ scheme. Finite subtractions are needed in order to restore helicity conservation at parton level \cite{CCFG}. 
A discussion of the longitudinal contributions to the real emissions, 
which are a subset of the total emissions considered here -and briefly analyzed below- can be found in \cite{CCFG}.

\section{ The qg sector and the Asymmetries}

When the incoming quark and the incoming gluon are both transversely 
polarized, the scattering matrix element is zero. 

On the contrary of the longitudinal case (LL scattering), in which the 
process takes contributions from the non-singlet and from the singlet sector, 
TT scattering in Drell Yan involves only the quark sector. 
 
This statement can be directly verified to lowest order by an inspection 
of the Dirac traces, and remains true to all orders in perturbation theory. 

To be definite, we introduce 2 (transverse) 
polarization vectors for the gluon, here denoted as $\epsilon_{1\perp}$ and 
$\epsilon_{2\perp}$, and choose a collinear base made out of the 2 momenta 
$p_1$ and $k_2$, the momenta of the initial state quark and of the final state quark or gluon. $p_2$ is the gluon momentum. All these partons are massless: 
$p_1=p_2^2=k_2^2=0$. 

The two transverse polarizations four-vectors are then defined as 
\beqa
&& {\epsilon_{1\perp}}_\mu={1\over \sqrt{2(k_1\cdot k_2)( p_1\cdot k_1) 
(p_1\cdot k_2)}} \left( (p_1\cdot k_1) {k_2}_\mu - (k_2\cdot k_1) 
{p_1}_\mu \right)
\nonumber \\  
&& {\epsilon_{2\perp}}_\mu={1\over \sqrt{2(k_1\cdot k_2) (p_1\cdot k_1) 
(p_1\cdot k_2)}}\epsilon_{\mu\alpha\beta\gamma} p_1^\alpha k_2^\beta k_1^\gamma .
\eeqa

One easily gets 
\beq
\not{\epsilon}_{2\perp}=(\not{p}_1\not{k}_2 \not{p}_2 - 
\not{p}_1 k_2\cdot p_2 + \not{k}_2 p_1\cdot p_2 - \not{p}_2 p_1\cdot k_2) 
\gamma_5 
\eeq
which can be easily implemented in symbolic calculations. 
With these definitions then $\epsilon_\perp\cdot k_1=\epsilon_\perp\cdot p_2=\epsilon_\perp\cdot k_2=0$, which guarantees transversality of the (physical) 
gluon. 

For initial state asymmetries, in order to have a nonvanishing trace, each 
spinor projector has to contribute with a zero or an even number of gamma matrices. The contribution $\sim \gamma_5\not{s}$ is clearly odd and therefore 
renders the total number of gamma matrices odd (vanishing) in each trace. 

A 
longitudinal quark, however, can scatter from a  transverse gluon, since only 
the $\sim\gamma_5\lambda$ part of the fermion projector becomes relevant. 
These contributions are part of the LL scattering cross section, since  
we can relate transverse gluon polarizations to the usual 
gluon helicities by a linear combination $\epsilon^\pm=1/\sqrt{2}( \epsilon_{1\perp} 
\pm i\epsilon_{2\perp})$. The vanishing of $A_{TT}$ in the qg sector indicates that the ratio between $A_{LL}$ and $A_{TT}$ is 1 to lowest order. 
This simple result shows that Drell Yan plays a key role in the analysis of the transverse spin distributions, since the gluon contributions is absent to all orders.

\section{The treatment of the real emissions}
In this section we focus our attention on the analysis of the real emissions. 
The steps that we describe here are necessary in order to proceed with  
an analytical or a numerical study of the real diagrams.

There are 2 basic modifications to be considered in this case: 1) the presence of a 2-to-4 final phase space integration; 2) the presence of transverse spin 
in the HV regularization. Some of the features appear already 
in the longitudinal case, but we prefer to present a comprehensive 
discussion of all these aspects for future reference. 

Let $v_1$ and $v_2$ denote the momenta of the 2 leptons in the final state.  
We also set $k_1=q$ 
for notational convenience and introduce the invariants
\beqa
&& s=(p_1 + p_2)^2,\nonumber \\
&& s_{12}=(k_1 + k_2)^2, \,\,\,\, s_{23}=(k_2 +k_3)^2,\nonumber \\
&& t_1=(p_1-k_1)^2, \,\,\, u_1=(p_2-k_1)^2, \nonumber \\
&& s_{13}=(k_1 + k_3)^2, \nonumber \\
&& t_2=(p_1-k_2)^2, \,\,\,t_3=(p_1-k_3)^2, \nonumber \\
&&  u_2=(p_2-k_2)^2, \,\,\,\,\, u_3=(p_2-k_3)^2, \nonumber \\
&&  {t'}_i=(p_1-v_i)^2, \,\,\,\,\, {u'}_i=(p_2-v_i)^2.
\eeqa
Only five of them are independent since 
\beqa
&& s=Q^2 -(u_1 +u_2 +u_3)= Q^2-(t_1 + t_2 +t_3 )=s_{12} +s_{23} + 
s_{13} -2 Q^2,
\nonumber \\
&& s_{12}=s + u_3 + t_3, \nonumber \\
&& s_{23}=s + u_1 + t_1 - 2 Q^2, \nonumber \\
&& s_{13}=s + u_2 + t_2, \nonumber \\
&& t_1= Q^2 + {t'}_1 + t'_2, \nonumber \\
&& u_1= Q^2 + u'_1 + u'_2. 
\eeqa
The integration of the 2-to-4 matrix elements introduces  
-beside the usual matrix elements of the LL cross section- new matrix 
elements containing ${t'}_1$, ${t'}_2$, ${u'}_1$ and ${u'}_2$. 

In the separation of the 2-to-4 into 2-to-2 subintegrals, we choose to work 
in two separate frames 1) the center-of-mass frame of the $k_2-k_3$ pair; 2)
the center-of-mass frame of the photon with axis fixed by the Collins-Soper 
\cite{CS} choice.

To be specific, we start from the 4-particle phase space integral is given by the formula

\beqa
PS_4\equiv\! \int\! {d^n v_1}d^n v_2\delta_+(v_1^2)\delta_+(v_2^2) 
{d^n k_3}{ d^n k_2}
 \delta(k_3^2)\delta(k_2^2)\delta^n(p_1 +p_2 - k_3 -k_2 -v_1-v_2).
\label{ps1}
\eeqa

We lump together the momenta $k_3$ and $k_2$  in $k_{23}$ and 
$v_1$ and $v_2$ in $q\equiv k_1$ as follows

\beqa
&& PS_4=\int d^n q\, d^n k_{23}
\delta^n(p_1 + p_2 - q - k_{23})
 \int d^n v_1\delta_+(v_1^2)\delta_+((q-v_1)^2) \nonumber \\
&& \times \int d^n k_2 \delta_+(k_2^2)\delta_+((k_{23}-k_2)^2).
\eeqa 
{}From this expression it is obvious that in the rest frame of the $(2,3)$ pair 
($k_{23}$ rest frame) the collinear singularities linked to the two gluons can be easily 
isolated in Dimensional Regularization. 

If hat-momenta are not present, then we easily get 

\beqa 
&& PS_4=\int d^n q \int d^n v_1 \delta_+(v_1^2)\delta_+((q- v_1)^2)
\int d^n k_{23}\delta^n(p_1 + p_2- q - k_{23})\nonumber \\
&& \hspace{2cm} \times \left(k_{23}^2\right)^{n/2-2}{\pi^{n/2-2-3/2}\over 
2^{n-2} \Gamma[n/2-3/2]}
\times I_{12}
\eeqa

where 
\beq
I_{1,2}=\int_{0}^{\pi}d\theta_1 \sin^{n-3}\theta_1
\int_{0}^{\pi}d\theta_2 \sin^{n-4}\theta_2
\eeq 
is evaluated in the rest frame of $k_{23}$ and from which 
the singularities can be exposed. 

The final result can be cast in the form 
\beq
 PS_4=\int d^n q \int_{\gamma}d^n v_1\delta_+(v_1^2) \delta_+((q- v_1)^2) \,
s_{23}^{n/2-2}{\pi^{n/2-3/2}\over 2^{n-2} \Gamma[n/2-3/2]}\times I_{1,2},
\label{psps}
\eeq

where we have used the relation $s+t + u= s_{23} + Q^2$. The remaining left-over integral has to be evaluated in the photon rest frame, with 
the Collins-Soper choice of axis. 
Finally (\ref{psps}) can be expressed in the form

\beqa
&& PS_4=\int d^n q\, d\Omega\,\Omega^{n-4}(Q^2)^{n/2-2}
 s_{23}^{n/2-2}{\pi^{n/2-3/2}\over 2^{n-2} \Gamma[n/2-3/2]}\times I_{1,2}.
\eeqa
Matrix elements containing ${t'}_1$ , such as ${t'}_1/(t_2 u_2)$ and similar, 
can all be exactly evaluated by this procedure, slightly generalized to the tensor case. Matrix elements containing $k_2\cdot v_1$ (and any denominator) 
are also easily handled in these two frames.  
After integration, the result can be conveniently expressed in the 
Collins-Soper (CS) frame. 
For this purpose \cite{CS} we parameterize the momentum of 
the photon in a collinear basis defined by the 
two incoming partons $p_1$ and $p_2$, and expand $q= x_1 p_1 + x_2 p_2 + 
q_T$. We introduce two metric projectors $g_T$ and $g_q$ defined by 

\beqa
&& g_T^{\mu\nu}=g^{\mu\nu}-{ p_1^\mu p_2^\nu\over p_1\cdot p_2} - 
{p_1^\nu p_2^\mu\over p_1\cdot p_2}\nonumber \\
&& g_q^{\mu\nu}=g^{\mu\nu} - {q^\mu q^\nu\over q^2}.
\eeqa

These two metrics 
project in the directions orthogonal to the collinear basis and to q respectively. 

An orthonormal set of 4-vectors $(\hat{q},\hat{z},\hat{x},\hat{y})$ 
is then defined as follows 
\beqa
\hat{q}={q^\mu\over Q},\,\,\,\,\,\,\,\,\,\,\hat{z}^\mu={Z^\mu\over \sqrt{-Z^2}},
\,\,\,\,\,\,\,\,\,\, \hat{x}^\mu={X^\mu\over \sqrt{-X^2}},\,\,\,\,\,\,\,\,\,\,\,
\hat{y}^\mu={Y^\mu\over \sqrt{-Y^2}},
\eeqa
where
\beqa
&& Z^\mu=\left({p_2\cdot q\over p_1\cdot p_2}\right)p_{q,1}^\mu -
\left({p_1\cdot q\over p_1\cdot p_2}\right)p_{q,2}^\mu,\nonumber \\
&& X^\mu=-\left({p_2\cdot Z\over p_1\cdot p_2}\right)p_{q,1}^\mu + 
\left({p_1\cdot Z\over p_1\cdot p_2}\right)p_{q,2}^\mu,\nonumber \\
&& Y^\mu={1\over p_1\cdot p_2}\epsilon^{\mu\nu\rho\sigma}p_{1\nu}
 p_{2\rho}q_{\sigma},\nonumber \\
&& p_{q,i}^\mu \equiv p_i^\mu- {p_i\cdot q\over q^2}q^\mu.
\eeqa

The set is orthogonal and spacelike. In the photon rest frame, therefore, 
it defines a cartesian set of axis. Note that in the collinear frame of the 
two incoming parton, $Z$ is collinear to $p_1$ and $p_2$, while $g_T$ projects over the transverse (w.r.t. $p_1$ and $p_2$) plane.

The vector $v_1$ can be expanded as 

\beq
v_1^\mu={ q^\mu\over 2} +{Q\over 2}\left( \sin\theta \cos\phi\, \hat{x}^\mu +
\sin\theta\sin\phi\, \hat{y}^\mu +\cos\theta\, \hat{z}^\mu\right)
\eeq 
and the cross section expressed in a form closer to eq.~(\ref{bebe}).
Finally, one expresses the scalar products $p_1\cdot v_1$, $q\cdot v_1$ 
or $s_{\perp i}\cdot v_1$ in the CS frame.

\section{Hat-momenta Integration}

As we have mentioned before, the HV prescription introduces matrix elements containing hat-momenta, due to separation of the n-dimensional space into 
4 and $n-4$ dimensions. 
 Beside $v_1$, 
the only other hat-momenta expected from the matrix elements are those of 
$k_2$ and $k_3$, which are treated in the way discussed below.  
Other hat-momenta can be eliminated by a convenient choice of axis. 
This causes, at least in part, a modification of the phase space integral which appear in the unpolarized case. 

In the case of unpolarized scattering,
 the singularities are generated by poles 
in the matrix elements which have the form $1/t_3$, $1/u_3$, $1/(t_3 u_3)$ and 
similar ones, in multiple combinations of them. Multiple poles can be reduced 
to sums of combinations of double poles by using simple identities  
among all the invariants and by the repeated use of partial fractioning. 
This is by now a well established procedure. 
In our case we encounter new terms of the 
form $1/t_3^2$ and $1/u_3^2$ and new matrix elements 
containing typical factors 
of the form ${\widehat{{k}}}_3$,  
$\widehat{{k}}_2$, and $\widehat{{k_2\cdot k_3}}$
at the numerator. In the longitudinal case, 
the hat-momenta that appears in the matrix elements are those related to
 $p_1,p_2, k_1$ and $k_2$. As we discuss in the Appendix, 
we can set to zero all the matrix elements 
containing hat-momenta of $p_1, p_2$ and $k_1$ by a convenient choice of axis. 
 Since the separation of the n-dimensional space 
into a  4 and an n-4 dimensional subspace is arbitrary, we can always assume that all the momenta which are not integrated over are embedded in the 
4-dimensional part. Therefore, 
in the case of a general spin dependent cross section, 
we can remove all the hat-momenta related to $p_1,p_2,s_{1\perp}$ 
and $s_{2\perp}$. The left-over hat-momenta, which are not set to vanish, 
are those related to $v_1,k_2$ and $k_3$, while $\hat{v_2}$ is reexpressed  
in terms of $\hat{q}$ and $\hat{v}_1$. Since $q$ is not integrated over, 
$\hat{q}$ is also zero. There are 2 ways to integrate the hat-momenta contributions. The most direct one is illustrated below. For instance, in the integration of $\hat{k}_2^2$, we use the on shell condition $k_2^2=0$ to relate 
$\hat{k_2}^2$ to the 4-dimensional projection $\widehat{\widehat{k_2}}$.

We start from the matrix element in the $(2,3)$ rest frame 
\beq
PS_2\equiv \int d^n k_2\delta(k_2^2)\delta((k_{23}- k_2)^2)
\widehat{\widehat{k}}_2^2
\label{trickp3}.
\eeq
We have set $k_2=(\widehat{\widehat{k}}_2,\widehat{k}_2)$, with
\beq 
\widehat{\widehat{k}}_2=k_2^0(1,\cos\theta_3 \sin\theta_2 \sin\theta_1,
\cos\theta_2 \sin\,\theta_1,\cos\,\theta_1)
\eeq

being the 4-dimensional part of $k_2$. 
We easily get 
\beq
\widehat{\widehat{k}}^2_2= {s_{23}\over 4}
\sin^2\theta_3\,\sin^2\theta_1\,\sin^2\theta_2.
\eeq

Therefore the usual angular integration measure 
\beq
d\Omega^{(n-2)}=\prod_{l=1}^{n-2}\,\sin^{n-l-2}\theta_l\,d\theta_l 
\eeq
with
\beq
\Omega^{(n-2)}=2 \prod_{l=1}^{n-2}\,\int_0^\pi \sin^{n-l-2}\theta_l\,d\theta_l 
\eeq
 
 is effectively modified to 
\beq
d\Omega^{(n-2)}=\prod_{l=1}^{3}\sin^{n-l}\theta_1 d\theta_l\times 
\prod^{n-2}_{l=4} \sin^{n-l-2}\theta_l d\theta_l. 
\eeq
The intermediate steps of the evaluation are similar to those in the previous section.

In $n=4-2 \epsilon$ dimensions we get
\beq
PS_3={ \pi^{2\epsilon} \epsilon\over 2^8 \pi^4 \Gamma[1-\epsilon]}
\left( {u t - Q^2 s_{23}\over s}\right)^{-\epsilon}
{s_{23}^{1-\epsilon}\over 2} \int_0^{\pi}d\theta_1 \,
\sin\theta_1^{3- 2\epsilon} \,\int_0^\pi \,d\theta_2 \, sin\theta_2^{2-2\epsilon},
\eeq

where $\theta_1$ and $\theta_2$ are the only relevant 
angles which appear in the matrix elements and therefore are not integrated. 
We have displayed also the $\theta_3$ integral since it is different from the 
unpolarized case.

The contributions from hat-momenta are either 
finite or of order $1/\epsilon$ and therefore play a key role in the 
cancelation of the mass singularities in the cross section. 

\section{Collinear Subtractions for $d\sigma_{NS}$}
The real emission diagrams (sets G,H,F) have collinear singularities which 
are generated by the emission of a massless parton off the initial state quark or gluon. Factorization in the quark sector involves diagrams from all the 3 sets mentioned above. As discussed in \cite{EMP} for the unpolarized case 
and in \cite{CCFG} in the case of longitudinal polarizations, the diagrams which require explicit factorization, in the non singlet sector, are those of the 
set G. The remaining collinear singularities cancel in the other real sets 
after adding all the contributions.  For this purpose we start defining 
the relation between the bare and the renormalized 
structure functions by the equation 

\beq
G_{A\to i}(x,M^2)= \int_x^1 {d\,z\over z}\left[ \delta_{i j}\delta(z-1) +
{\alpha_s\over 2 \pi}R_{i\leftarrow j}(z,M^2)\right] G^{bare}_{A\to i}
\left({x\over z}\right).
\eeq
with $R$ of the form 
\beq
R_{i\leftarrow j}= -{1\over \epsilon}\Delta P_{ji }(z)
{\Gamma[1-\epsilon]\over \Gamma[1-2\epsilon]}\left({4 \pi \mu^2\over M^2}
\right)^{\epsilon} + C_{i\leftarrow j}(z)
\eeq
for the LL subtractions and 
\beq
R_{T q\leftarrow q}= -{1\over \epsilon}\Delta_T P_{qq}(z)
{\Gamma[1-\epsilon]\over \Gamma[1-2\epsilon]}\left({4 \pi \mu^2\over M^2}
\right)^{\epsilon} + C_{q\leftarrow q}(z)
\eeq
for the TT (or transverse spin) subtractions. 
where the finite pieces $C_{i\leftarrow j}(z)$ are arbitrary. 
M is the factorization scale. 

In the
$\overline{MS}$ 
scheme they are set to be zero. The longitudinal, $\Delta P_{ij}$, 
and the transverse, $\Delta_TP_{qq}$, splitting  functions are given by 
\beqa
&&\Delta P_{qq}= P_{qq}=C_F\left[ {(1+z^2)\over (1-z)_+} + \frac{3}{2} \delta(1-z)\right] \nonumber \\
&&\Delta P_{qg}=\frac{N_F}{2}[ 2 z-1]\nonumber\\
&& \Delta P_{gg}=N_C\left[ (1+z^4)\left(
\frac{1}{z}+\frac{1}{(1-z)_+}\right)-\frac{(1-z)^3}{z}
\right]+\frac{33-2 N_F}{6}\delta(1-z)\nonumber \\
&&\Delta P_{gq}=C_F[2-z],\nonumber \\
&& \Delta_T P_{ qq}=C_F\left( {2 x\over (1-x)_+} +{3\over 2}\delta(1-x)\right)
\eeqa

with the distribution $1/(1-z)_+$ defined by 
\beq
\int_0^1 d\, z{ f(z)\over (1-z)_+}=\int_0^1{f(z)-f(1)\over 1-z}.
\eeq

Beside the usual subtractions of the singularities in the 
longitudinal spin sector, now we have also 
similar subtractions in the transverse spin sector. 

As we have mentioned above, the collinear subtractions, in $d\sigma_{NS}$ 
involve only the set G. 
In these diagrams a quark (or an antiquark) 
can emit a collinear gluon from the initial state. $R_{i\leftarrow j}$ is therefore proportional to $\Delta P_{qq}$  $(=\Delta P_{\bar{q}\bar{q}})$ or to 
$\Delta_T P_{ qq}$. 

The non singlet cross section is given to order 
$\alpha_s^2$ by the renormalized cross section $d\hat\sigma_1$ with the 
the collinear initial and final state singularities subtracted
\begin{eqnarray}
s {d\hat\sigma^{fact}\over d\,t d\,u} &=& 
s {d{\hat\sigma_1}\over d\,t d\,u}\nonumber \\
 & - & {\alpha_s\over 2 \pi} 
\int_0^1 d\,z_1 R_{q\leftarrow q}(z_1,M^2)s {d\sigma^{(1)LL}_{qq}\over
d\, t}\mid_{p_1\to z_1 p_1}
\delta\left(z_1(s + t - Q^2) + u\right)\nonumber \\
&-& {\alpha_s\over 2 \pi} \sum_k
\int_0^1 d\,z_2 R_{q\leftarrow q}(z_2,M^2)s {d\sigma^{(1)LL}_{qq}\over d\,
t}\mid_{p_2\to z_2 p_2}
\delta(z_2(s + u - Q^2) + t)\nonumber \\
& - & {\alpha_s\over 2 \pi} 
\int_0^1 d\,z_1 R_{T q\leftarrow q}(z_1,M^2)s {d\sigma^{(1)TT}_{qq}\over
d\, t}\mid_{p_1\to z_1 p_1}
\delta\left(z_1(s + t - Q^2) + u\right)\nonumber \\
&-& {\alpha_s\over 2 \pi} \sum_k
\int_0^1 d\,z_2 R_{q\leftarrow q}(z_2,M^2)s {d\sigma^{(1)TT}_{qq}\over d\,
t}\mid_{p_2\to z_2 p_2}
\delta(z_2(s + u - Q^2) + t).
\label{fact}
\end{eqnarray}

In Eq.~(\ref{fact}) $d\sigma^{LL}$ and $d\sigma^{TT}$ are Born level 
cross sections at a rescaled value of one of the two incoming momenta (
$z_1 p_1$ or $z_1 p_2$ ) due to the collinear gluon emission, and evaluated 
in n dimensions.

 LL and TT refer to their longitudinal/transverse spin content and are proportional to $T_1$ and $T_2$ respectively.

\section{Conclusions}

We have presented a general discussion of the spin dependence 
of the Drell Yan cross section for the nonzero $p_T$ distributions, 
and shown that the analysis of the hard scattering can be performed in the 
t'Hooft-Veltman scheme consistently. 
Factorized expressions for the virtual corrections have been presented. 
Our analysis has been focused on the non singlet sector, which is the main source of transverse spin dependence in this process. The result for the 
virtual corrections presented here, however, are those of the entire process. 

{}From our results, we find that -at non zero $p_T$- a transverse spin 
dependence is already generated from the diagonal part of the leptonic 
tensor. The calculation is an explicit proof of factorization of the 
hard scattering , 
in a nontrivial case.
We have, along the way, 
developed a complete methodology for the analysis of the real emissions in 
the same scheme. 

We should also mention that factorization theorems for 
spin-dependent cross sections are yet to be proved. 
In fact, all the results presented in the literature are obtained 
assuming a light-cone dominance of the process. 
 Particularly important -and unsettled- 
appears the issue of higher twist effects in the factorization formula 
and in the expression of the $A_{LT}$ asymmetries.

Work toward 
a complete analysis of the hard scattering for the nonzero $p_T$ distributions 
is now in progress and we hope to return on some of these 
issues in a future work.

\centerline{\bf Acknowledgements}
 
We are very grateful to John Collins 
and to George Sterman for patient discussions and suggestions. 
C.C. thanks L. Gordon, Hsiang-nan Li and 
Janwei Qiu for discussions and the ITP at Stony Brook for hospitality. 
We have used {\em FeynCalc} \cite{Feyn} and {\em Tracer} 
\cite{tracer} in the symbolic calculations.

\section{Appendix. The 2-particle phase space}

Let's start considering the virtual contributions to the cross section.
The relevant phase space integral is given by
\begin{equation}
PS_2=\int \frac{d^n q\, d^nk_2}{(2\pi)^{n-1}(2\pi)^{n-1}}\delta_+(q^2-Q^2)
\delta_+(k_2^2)\delta^n(p_1+p_2-k_2-q) (2\pi)^n
\end{equation}
It is convenient to introduce light-cone variables
$(q^+,q^-,q_\perp)$, with
\begin{equation}
q=q^+n^+ + q^-n^- + q_\perp
\end{equation}
and
\begin{equation}
n^\pm\equiv \frac{1}{\sqrt{2}}(1, 0_\perp, \pm 1)
\end{equation}
and work in the frame
\begin{equation}
p_1^+=p_2^-=\sqrt{\frac{s}{2}}
\end{equation}
to get
\begin{eqnarray}
PS_2 &=  & \int \frac{d^+qd^-q}{(2\pi)^{n-2}}|q_\perp^2|^{n/2-2} d|q_\perp^2|
\delta(2q^+q^- -|q_\perp^2|-Q^2)\delta(s+t+u-Q^2)\\
 & = &  \int d^+qd^-q(2q^+q^- Q^2)^{n/2-2}\delta(s+t+u-Q^2)
\frac{\Omega^{n-3}}{(2\pi)^{n-2}}
\end{eqnarray}
with 
\begin{equation}
\Omega^{n-3}\equiv 2\int^\pi_0 \prod^{n-3}_{l=1} \sin\theta_l^{n-l-3} d\theta_l
=\frac{2 \pi^{n/2-1}}{\Gamma(n/2-1)}
\end{equation}
using
\begin{equation}
\frac{\partial(q^+,q^-)}{\partial(t,u)}=\frac{1}{2s},
\end{equation}
and after covariantization
\begin{equation}
(2q^+q^- -Q^2)\rightarrow \frac{(Q^2-t)(Q^2-u) -s Q^2}{s} =\frac{ut}{s},
\end{equation}
\begin{equation}
PS_2 =\int dt du \frac{1}{2s}\left(\frac{ut}{s}\right)^{n/2-2}
\frac{2 \pi^{n/2-1}}{\Gamma(n/2-1)}
\frac{\delta(s+t+u-Q^2)}{(2\pi)^{n-2}},
\end{equation}
\begin{equation}
\sigma= \frac{1}{4 N_c}\int \frac{dtdu}{(2s)^2} 4\pi \alpha_s 
\left(\frac{\mu^2}{Q^2}\right)^\epsilon
\left(\frac{sQ^2}{ut}\right)^\epsilon \delta(s+t+u-Q^2)T_0(Q^2,u,t).
\end{equation}

The integration to the total cross section can be obtained as follows. 
We start from 
\beq
\int d PS_2 = {1\over \pi^{n/2-1} 2^{2 n -4} \Gamma[n/2-1]}
\left({s - Q^2\over s}\right)^{n-3} s^{n/2-2}\int_0^\pi
\, (\sin\theta)^{n-3} d\theta.
\label{ps2}
\eeq

We also define 
\beq
J=\int_{0}^{\pi} d\theta \sin \theta^{n-3}d\,\theta.
\eeq
We define as usual 

\beq
s = (p_1 + p_2)^2\,\,\, t=(p_1-k_1)^2 \,\,\, u=(p_2-k_1)^2 \,\,\, k_1^2= Q^2
\eeq
with $ s + t + u = Q^2$ and $Q^2$ is the invariant mass of the photon
 
We define the Bjorken variable $x=Q^2/s$  ($0<x<1$) 
and we assume that $x$ is not parametrically small ($x\sim 0$) or close to 1, $x\sim 1$. Both the small-x region 
and the large-$x$ $(x\to 1)$ need resummation, since large logs in the 
variables $x$ and $1-x$ are generated.
The scattering angle is $\cos \theta= {(Q^2 - 2 u - s)/ (s - Q^2)}$ and $u$ 
varies in between $-(s - Q^2) < u< 0$,  (i. e. $-s (1-x) <u<0$). 

It is convenient to introduce the angular variable $y= 1/2(1 + \cos\theta)$
and rewrite (\ref{ps2}) as 

\beq
\int PS_2= { \pi^{n/2-1}\over 2 \Gamma[n/2-1]} 
\left( {s - Q^2\over s}\right)^{n-3}s^{n/2-2}
\int_0^1 d\, [y (1-y)]^{n/2-2}. 
\eeq

We get 
\beq
J=2^{n-3}\int_{0}^{1} d\,y (y (1-y))^{n/2-2}.
\eeq

In terms pf the variable $y$ we have $u=-Q^2/x (1-x) y$ and 
$t=-Q^2/x (1-x)(1-y)$.

We set  $n= 4 + 2 \epsilon$ and define
\beq
K[n,s]={\pi^{n/2-1}\over 2 \Gamma[n/2-1]}\left({s - Q^2\over s}\right)^{n-3}
s^{n/2-2}.
\eeq

Using these notations $PS_2[\,,\,]= K[n,s] J[\,,\,] $. 

We get
\beqa
&& PS_2[1/(u t)]= K[n,s]{ x^2\over (Q^2)^2 (1-x)^2}
{\Gamma[n/2-2]^2 \over \Gamma[n-4]},\nonumber \\
&& PS_2[1/t]= PS_2[1/u]={- x\over (1-x) Q^2}K[n,s]
{\Gamma[n/2-2] \Gamma[n/2-1] \over \Gamma[n-3]}, \nonumber \\
&& PS_2[u/t]=PS_2[t/u]= K[n,s] {\Gamma[n/2-2]\Gamma[n/2]\over \Gamma[n-2]},
\eeqa
which can be expanded up to the desired order in $\epsilon$. 
For instance, the real emissions from the $q\bar{q}\to\gamma^*$ sector 
give 
\begin{eqnarray}
\hat\sigma = \left({2\pi e^2e_f^2 \over s}\right)
\left({1-\eps \over 2N_C}\right)\left({4\pi \mu^2 \over Q^2}\right)^{\eps}
\left({\alpha_s C_F \over 4\pi}\right)\left({\Gamma(1-\eps) \over
\Gamma(1-2\eps)}\right) \nonumber \\
\mbox{}\left[(1-h)\left({4 \delta (1-x) \over \eps^2}
-{4 \over \eps} {1+x^2 \over (1-x)_+}
+8(1+x^2)\left({\rm{ln}(1-x) \over (1-x)}\right)_+ \right. 
\right. \nonumber \\ 
\left. \mbox{}-4\rm{ln}(x){1+x^2 \over (1-x)_+}\right) 
+16h\left({-\delta(1-x) \over
2 \eps}-{\delta(1-x) \over 2}+{1-x+x^2 \over (1-x)_+}\right) \nonumber \\
\mbox{}+(s_{1\perp} \cdot s_{2\perp})\left({-4\delta(1-x) \over \eps} +
{2(1-x) \over \eps}+{2(3-2x+3x^2) \over (1-x)_+} \right. \nonumber \\
\left. \left. \mbox{}-4\delta(1-x)
+2(1-x)\rm{ln}\left({x \over (1-x)^2}\right)\right)\right].
\label{crosss} 
\end{eqnarray}
We have factorized the expression in this manner to facilitate
comparison with the literature \cite{aem}, in which the normalization
convention is to remove the first two quantities in parentheses in 
Eq.~(\ref{crosss}).

\section{Appendix. Integration of the real emission diagrams}

In order to integrate over the matrix elements, we need to evaluate the 
various scalar products which appear in such matrix elements,  in the 
c.m. frame of the pair $(1,2)$. For this purpose we define the 
functions 

\beqa
&&\lambda(x,y,z)= x^2 + y^2 +z^2-2 x y -2 y z -2 x z \nonumber \\
&& P[x,y,z]={\lambda^{1/2}(x,y,z)\over 2 \sqrt{x}} \nonumber \\
&& E[x,y,z] ={x+y-z\over 2 \sqrt{x}}.
\eeqa
It is easy to show that 

\beqa
&& |\vec{p}_1|=P[s_{23}, p_1^2,u_1]={s_{23}- u_1\over 2 \sqrt{s_{23}}},
\nonumber \\
&& |\vec{p}_2|=P[s_{23},p_2^2,t_1]=
{s_{23}- t_1\over 2 \sqrt{s_{23}}}, \nonumber \\
&&|\vec{k}_3|=|\vec{k}_2|=P[s_{23},p_1^2,p_2^2]=
{\sqrt{s_{23}}\over 2},\nonumber \\
&& |\vec{k}_1|=\sqrt{s\over s_{23}}P[s,Q^2,s_{23}]=
{\sqrt{\lambda(s, Q^2,s_{23})}\over 2 \sqrt{s_{23}}},\nonumber \\
&& p_1^0=E[s_{23},p_1^2,u_1]={s_{23}- u_1\over 2 \sqrt{s_{23}}},\nonumber \\
&& p_2^0=E[s_{23},p_2^2,t_1]={s_{23}- t_1\over 2 \sqrt{s_{23}}}.
\label{list}
\eeqa

We define
\beqa
&& v = 1 + {t- Q^2\over s}, \nonumber \\
&& w= - {u\over s + t - Q^2}.
\eeqa
We get
\beqa
&& p_1^0={s_{23}-u_1\over 2 \sqrt{s_{23}}}=
 {s v\over 2 \sqrt{s_{23}}},\nonumber \\
&& p_2^0={s_{23}-t_1\over 2 \sqrt{s_{23}}}=
{s(1-v) - Q^2\over 2 \sqrt{s_{23}}},\nonumber \\
&& k_1^0={Q^2 -s + s_{23}\over 2 \sqrt{s_{23}}}=
{s (1-v + v w) - Q^2\over s \sqrt{s_{23}}},\nonumber \\
&& k_3^0=k_2^0={\sqrt{s_{23}}\over 2}.
\label{k3}
\eeqa

In the derivation of (\ref{k3}) we have used the relation 
$s+ t_1 + u_1= Q^2 + s_{23}$. 
There are four different parameterizations of the integration momenta 
which we will be using. In the first one, which is suitable for unpolarized 
scattering one defines (in the c.m. frame of the pair $(1,2)$)
in general 

\beqa
&& k_3={\sqrt{s_{23}}\over 2}(1,...,\cos\theta_2 \sin\theta_1,\cos\theta_1),
\nonumber \\
&& k_2={\sqrt{s_{23}}\over 2}(1,...,-\cos\theta_2 \sin\theta_1,-\cos\theta_1),
\nonumber \\
&& p_1=p_1^0(1,0,...,0,\sin\psi_1,\cos\psi_1),\nonumber \\
&& p_2=p_2^0(1,0,...0,\sin\psi_1,\cos\psi_1),\nonumber \\
&& k_1=(k_1^0,0,...0,|k_1|\sin \psi_2, |k_1|\cos \psi_2),
\label{kk}
\eeqa
where the dots denote the remaining $n-2$ polar components. 
We get 
\beqa
&& \sin\psi_1=\left( {s (1-w)\over s (1- v w) - Q^2}\right)^{1/2},
\nonumber \\
&& \sin \psi_2={\sin \psi (Q^2 -s (1-v - v w))\over \lambda^{1/2}(s, Q^2, 
s_{23})}.
\eeqa
It is convenient to use the parameterizations 
\begin{itemize}
\item{set 1}
\beqa
&& p_1=p_1^0(1,0,...,0,0,1),\nonumber \\
&& p_2=p_2^0(1,0,...,-\sin\psi'',0,\cos\psi''),\nonumber \\
&& k_1=(k_1^0,0,...,-|\vec{k}_1|\sin\psi,0,|\vec{k}_1|\cos\psi),
\eeqa

\item{set 2}
\beqa
&& p_1=p_1^0(1,0,...,\sin\psi'',0,\cos\psi''),\nonumber \\
&& p_2=p_2^0(1,0,...,0,0,1),\nonumber \\
&& k_1=(k_1^0,0,...,|\vec{k}_1|\sin\psi',0,|\vec{k}_1|\cos\psi'),
\eeqa

\item{set 3}
\beqa
&& p_1=p_1^0(1,0,...,\sin\psi,0,\cos\psi),\nonumber \\
&& p_2=p_2^0((1,0,...,-\sin\psi',0,\cos\psi'),\nonumber \\
&& k_1=(k_1^0,0,...,0,0,|\vec{k}_1|),
\eeqa
\end{itemize}
where $0,...$ refers to $n-5$ components identically zero. 
It is straightforward to obtain the relations 

\beqa
&& \cos\psi''={(s_{23}-t_1)(s_{23}- u_1) - 2 s_{23} s\over (s_{23}- t_1)(s_{23}- u_1)},\nonumber \\
&& \cos\psi={(-Q^2 +s - s_{23})(s_{23}- u_1)-2 s_{23}(Q^2 - t_1)\over 
\lambda^{1/2}(s,Q^2,s_{23})(s_{23}- u_1)},
\nonumber \\
&& \cos\psi'={(s_{23}- t_1)(-Q^2 +  - s_{23}) -2 s_{23}(Q^2 - u_1)\over 
\lambda^{1/2}(s,Q^2,s_{23})(s_{23} - t_1)}.
\eeqa

We select a specific set depending upon the 
form of the hat-momenta in the matrix elements.
We choose the center of mass of the gluon pair 
$(2,3)$. 
Following the notation of reference \cite{bennaker,thomas} we define 

\beqa
I_n^{(k,l)}&\equiv& \int_0^\pi d\,\theta_1\,\int_0^\pi\, d\,\theta_2 
\, \sin\,\theta_2^{n-4}\left(a + b \cos\,\theta_1\right)^{-k}\nonumber \\
&&\times
 \left(A + B \cos\,\theta_1 + C \sin\,\theta_1 \cos\,\theta_2
\right)^{-l}, 
\label{ben}
\eeqa
where a,b, A, B, C, are functions of the external kinematic variables . 

This expression trivially contains collinear singularities if $k\geq 1,\,\, 
l\geq 1$ and $a^2=b^2$ and/or $A^2 = B^2 + C^2$. The singularities are traced back to the emission of massless gluons from the initial state and in the final 
states. In this special case, it is convenient to rescale the integral 
and let the angular variables $\psi,\, \psi',\, \psi''$ defined above appear. 
The cases 
$a^2=b^2$ and $A^2\neq B^2 + C^2$, $a^2\neq b^2$ and $A^2 = B^2 + C^2$
and $a^2\neq b^2$ and $A^2 \neq B^2 + C^2$ are discussed in 
ref.~\cite{bennaker}. It is easy to figure out from the structure of the matrix elements which cases require a four dimensional integration and which, 
instead, have to be evaluated in $n$ dimensions. This procedure is standard 
lore. Notice that only two independent angular variables 
$\theta_1$ and $\theta_2$ appear at the time.

Matrix elements containing more than 2 integration invariants 
(for instance $1/t_2 u_2 u_3$ or $1/t_2 t_3 u_2$) have to be partial 
fractioned by using the Mandelstam relations. 
Just to quote an example, 
we can reduce these ratios in a form suitable for integration 
in the $\theta_1$, $\theta_2$ variables by partial fractioning
\beq
{1\over t_2 u_2 u_3}= {1\over t_2 u_2 u_3}\left( {u_2 + u_3\over Q^2 - s - u_1}
\right)
\eeq
where we have used the identity 
\beq
{u_2 + u_3\over Q^2 -s - u_1}=1.
\eeq
If we let $I[.]$ denote the corresponding angular integral, we get 
\beq
I[1/t_2 u_2 u_3]= {1\over Q^2 -s - u_1}
\left(I[1/t_2 u_2] +I[1/t_2 u_3]\right).
\eeq
The integrals on the rhs of this equation are now in the standard 
$\theta_1$, $\theta_2$ form. 

Defining 
\beqa
 V_n[\mp,\mp]\equiv \int_{0}^{\pi}\int_{0}^{\pi}d\theta_1 
d\theta_2 {\sin^{n-3}\theta_1\sin^{n-4}\theta_2\over 
(1- \cos\theta_1)^i(1\mp \cos\chi \cos\theta_1 \mp \sin\chi\cos\theta_2\sin \theta_1)^j},
\eeqa
we get
\beqa
&& V_n[-,-]= C[n,i,j]F[1,j,n/2-1,\cos^2{\chi\over 2}],\nonumber \\
&& V_n[+,-]=C[n,i,j]F[1,j,n/2-1,1/2 +\sin{\chi\over 2}],\nonumber \\
&& V_n[-,+]= V_n[-,-],\nonumber \\
&& V_n[+,+]= C[n,i,j]F[1,j,n/2-1,\sin^2{\chi\over 2}],
\eeqa
where
\beqa
C[n,i,j]\equiv 2^{1-i -j}\pi{\Gamma[n/2-1-j]\Gamma[n/2-1-i]\over \Gamma[n-2-i-j]}
{\Gamma[n-3]\over \Gamma^2[n/2-1]}.
\eeqa

Use of the sets 1,2,3 and 4 allows us to get rid of a large part of the
hat-momenta involved in the real emissions.
As an example of integration over the hat-momenta, we consider the matrix
element

\beq
 I\left[{\hat{k}_2\cdot \hat{v}_2\over t_2 u_2}\right]=
\int d^n v_1\delta_+(v_1^2)\delta(_+(q - v_1)^2)\,
\int d^n k_2\delta_+(k_2^2)\delta_+((k_{23}-k_2)^2)
{\hat{k}_2\cdot v_1\over t_2 u_2},
\eeq
and the sub-integral in the (2,3) frame can be expanded covariantly
\beq
\int d^n k_2{k_2^\alpha\over t_2 u_2}
\delta_+(k_2^2)\delta_+((k_{23}- k_2)^2)=A k_{23}^\alpha + B p_1^\alpha + C
p_2^\alpha.
\eeq

The evaluation of the coefficients A B and C can be carried out by standard
methods. However, 
since we project the n-4 dimensional components by the metric
$\hat{g}^{\alpha\beta}$ and since $\hat{p}_1=\hat{p}_2=\hat{q}=0$, then we
easily find that 
the contribution of this matrix element is zero. Therefore, the
only hat-momenta left over are those involving $k_{2}^2$ and $v_1^2$
(i.e. $\hat{k}_2^2$ and $\hat{v}_1^2$) which can be treated as discussed in
section 13.

\section{Appendix. A fast way to renormalize}

In this appendix we illustrate a simple method developed by two of us 
\cite{CC} 
to obtain renormalized expressions for any Feynman diagram at one loop. The partons are set on-shell from the 
beginning and UV and IR divergences are identified. After performing the PV reduction, a large number of massless tadpoles beside self-energies, scalar triangles diagrams 
and scalar box diagrams, are generated. We identify a set of prescriptions for handling massless tadpoles and implement them symbolically. They can all be derived 
by renormalizing off-shell, i.e. by 
subtracting the UV divergence while keeping the external lines off mass 
shell. We omit the derivation of these rules, which have been applied 
before \cite{CCFG} and just describe here some simple applications. 
The renormalization of the other scalar diagrams is performed as usual, by subtracting the UV poles whenever they appear.  

\subsection{Self-energy diagram}
Let's start from the scalar self-energy contribution which is given by  
\begin{eqnarray}
I_1(p)&=&\int {d^n l\over (2 \pi)^n}{1\over l^2 (l + p)^2}\nonumber\\
&=&-i (4 \pi)^{-n/2}(-p^2)^{\omega/2}{2\over \omega} 
{\Gamma(1-\omega/2)\Gamma^2(1 +\omega/2)\over \Gamma(2 + \omega)},
\end{eqnarray}
where $n=4+2\omega$.
The PV function $B_0(p^2)$ is defined by
\begin{equation}
B_0(p^2)=\frac{(2\pi\mu)^{4-n}}{i\pi^2} I_1(p).
\end{equation}
In the PV reduction, isolated $B_0(0)$ terms are renormalized off-shell,  
to give
\begin{eqnarray}
B_0\to B_0^{\rm ren}(0)&=& \frac{1}{\omega},
\end{eqnarray}
while terms containing the product $\omega\times B(0)$  
are set to vanish
\begin{eqnarray}
\omega B_0(0)&=& 0.
\end{eqnarray}
For off-shell self-energy diagrams we proceed exactly in the same way.
We renormalize isolated self energy contributions which then take the form  
\begin{eqnarray}
B_0(p^2)&=& \left(\frac{-p^2}{\mu^2}\right)^\omega \left(2-\frac{1}{\omega} \right)
+\frac{1}{\omega}  + O(\omega)
\nonumber\\
&\simeq& 2- \ln\left(\frac{-p^2}{\mu^2}\right) + O(\omega)
\end{eqnarray}
but we leave unrenormalized all the expressions containing the self energy times an 
$\omega$ factor 
\begin{eqnarray}
\omega B_0(p^2)&=& -1 + O(\omega).
\end{eqnarray}
The latter result can be obtained by using the unrenormalized expression of the self-energy 
\begin{equation}
B_0^{\rm unren}(p^2)=\left(\frac{-p^2}{\mu^2}\right)^\omega \left(2-\frac{1}{\omega} \right)
\end{equation} 
instead of the renormalized one. 
We will see that these prescription allow us to reproduce all the 
one loop vertices needed in the calculation 
quite straightforwardly. It is also easy to check that 
the coefficients of the reduction of the (tensor) box diagrams are finite, 
as they should, although the PV reduction introduces spurious singularities. 

\subsection{Scalar triangular vertices}
The general scalar vertex integral is
\begin{equation}
I_2(p_1,p_2)=\int {d^n l\over (2 \pi)^n}{1\over l^2 (l + p_1)^2 (l + p_2)^2},
\end{equation}
and PV function is defined as
\begin{equation}
C_0(p_1^2,p_3^2,p_2^2)=\frac{(2\pi\mu)^{4-n}}{i\pi^2} I_2(p_1,p_2),
\end{equation}
where $p_3=p_1-p_2$.

In our paper, we need two different types of PV scalar vertex functions. 

\noindent{$\bullet$ \bf Case (1)} 
$p_1^2=p_3^2=0$ and $p_2^2=q^2\neq 0$.\\
$p_1$ is the momentum of the incoming quark, 
$p_3=p_1+p_2$ is the momentum of the gluon and
$p_2$ is the momentum of the virtual quark.
\begin{equation}
I_2 =-i (4 \pi)^{-n/2}(-{p_2}^2)^{-1 + \omega/2}{4\over \omega^2}
{ \Gamma(1-\omega/2)\Gamma^2(1 + \omega/2)\over \Gamma(1 + \omega)}.
\end{equation}
If we set $p_2^2=q^2$ the PV function can be reduced to
\begin{eqnarray}
C_0(0,0,q^2) &=& \frac{1}{q^2}\left(\frac{-q^2}{\mu^2}\right)^\omega
\left(\frac{1}{\omega^2}-\frac{\pi^2}{12} \right) + O(\omega)
\nonumber\\
&\simeq&  \frac{1}{q^2}\left(\frac{1}{\omega^2}
+ \frac{L_{q^2}}{\omega} +\frac{6L_{q^2}^2-\pi^2}{12}\right),
\end{eqnarray}
where $L_{q^2}\equiv\ln\left(\frac{-q^2}{\mu^2}\right)$.

\noindent{$\bullet$ \bf Case (2)}
$p_1^2=0$, $p_2^2=Q^2$ and $p_3^2=q^2$.

$p_1$ the momentum of the incoming quark, $p_2$ 
is the momentum of the photon and $p_3$ is the momentum of the virtual quark.
In this case
\begin{equation}
I_2 =-i (4 \pi)^{-n/2}{4\over \omega^2}
{\Gamma(1-\omega/2)\Gamma^2(1 + \omega/2)\over \Gamma(1 + \omega)}
{1\over p_2^2 - p_3^2} \left[ (-p_2^2)^{\omega/2} - 
(-p_3^2)^{\omega /2} \right]. \nonumber \\
\end{equation}
The PV function is
\begin{eqnarray}
C_0(0,q^2,Q^2) &=  &  \frac{1}{Q^2+q^2}\left\{\left(
\frac{-q^2}{\mu^2}\right)^\omega -
\left(\frac{Q^2}{\mu^2}\right)^\omega\right\}
\left(\frac{1}{\omega^2}-\frac{\pi^2}{12} \right) + O(\omega)
\nonumber\\
&\simeq&  \frac{1}{Q^2+q^2}\left(
\frac{L_{q^2}-L_{Q^2}}{\omega} +\frac{L_{q^2}^2-L_{Q^2}^2}{2}\right).
\end{eqnarray}

\subsection{gluon(photon)-quark-quark vertex (type 1)}
As an application of the renormalization procedure we illustrate here the 
derivation of the vertices which are needed in the calculation. 
Notice that if we use the method discussed above we don't have to  use any 
symmetric integration in order to isolate the coefficients of the scalar expansion.
\begin{equation}
V^\mu=\frac{(2\pi\mu)^{4-n}}{i\pi^2}\int d^n l \frac{\gamma^\alpha 
\!\!\not{l} \gamma^\mu (\!\not{l} + \!\!\not{p_2})\gamma^\alpha }{l^2(l+p_1)^2(l+p_2)^2}.
\label{eqv}
\end{equation}
By tensor reduction, we can set
%\begin{equation}
%V=\frac{(2\pi\mu)^{4-n}}{i\pi^2}\int d^n l \frac{(2-n) \!\!\not{l} \gamma^\mu %\!\!\not{l} + 
%(4-n)\!\!\not{l} \gamma^\mu\!\!\not{p_2}- 2\!\!\not{p_2}\gamma^\mu\!\!\not{l} %}{l^2(l+p_1)^2(l+p_2)^2}
%\end{equation}
\begin{equation}
\frac{(2\pi\mu)^{4-n}}{i\pi^2}\int d^n l \frac{l^\alpha l^\beta }{l^2(l+p_1)^2(l+p_2)^2}
=g^{\alpha\beta} C_{00} + p_1^\alpha p_1^\beta C_{11} + p_2^\alpha p_2^\beta C_{22}
+ (p_1^\alpha  p_2^\beta +p_2^\alpha p_1^\beta) C_{12}. 
\end{equation}
We also introduce the expansion
\begin{equation}
\frac{(2\pi\mu)^{4-n}}{i\pi^2}\int d^n l \frac{l^\alpha }{l^2(l+p_1)^2(l+p_2)^2}
= p_1^\alpha C_1 + p_2^\alpha C_2. 
\end{equation}
\noindent{\bf 1)} We take $p_1^2=0, p_2^2=Q^2$ and $(p_1-p_2)^2=0$. 
Then Eq.~(\ref{eqv}) becomes,
\begin{eqnarray}
V^\mu&=&(2-n)\left\{
\left((2-n) C_{00} - p_2^2 C_{22}-  \!\!\not{p_2}\!\!\not{p}_1 C_{12}\right)\gamma^\mu 
+2\!\!\not{p_2}p_2^\mu C_{22} +2\!\!\not{p_2}p_1^\mu C_{12}\right\} 
\nonumber\\
&&+(2-n)\left(2\!\!\not{p_2}p_2^\mu- p_2^2\gamma^\mu\right) C_2
-2\left(2\!\!\not{p_2}p_1^\mu-\!\!\not{p_2}\!\!\not{p}_1\gamma^\mu \right) C_1.
\end{eqnarray}
Notice that after applying the PV reduction we will encounter terms 
of the form $B_0(0)$ and $\omega B_0(0)$ which must be handled with the 
rules given above.
In fact we get for the tensor coefficients

\begin{eqnarray}
&& C_1 = 2{ B_0(0)\over Q^2} - 2{ B_0(Q^2)\over Q^2} - C_0(0,0,Q^2),\nonumber \\
&& C_2 = -{B_0(0)\over Q^2} + {B_0(Q^2)\over Q^2}, \nonumber \\
&& C_{00}= {1\over 4} + {B_0(Q^2)\over 4}, \nonumber \\
&& C_{11}= {1\over Q^2} - 3 {B_0(0)\over Q^2} 
+ 3 {B_0(Q^2)\over Q^2} + C_0(0,0,Q^2),\nonumber \\
&& C_{22} ={ B_0(0)\over 2 Q^2} - {B_0(Q^2)\over 2 Q^2}, \nonumber \\
&& C_{12} = -{1\over 2 Q^2} + {B_0(0)\over 2 Q^2} - 
{B_0(Q^2)\over 2 Q^2}.\nonumber \\
\end{eqnarray}

The vertex correction becomes,
\begin{eqnarray}
V^\mu&=& (2-n)
\left((2-n) C_{00} - Q^2 C_{22}-  Q^2 C_{12}\right)\gamma^\mu 
+(n-2) Q^2\gamma^\mu C_2
+2Q^2\gamma^\mu C_1 \nonumber\\
&=& \gamma^{\mu}\left(-\frac{2}{\omega^2} - \frac{2L_{Q^2} -4}{\omega} -8
+\frac{\pi^2}{6} +3L_{Q^2} -L_{Q^2} ^2 \right).
\end{eqnarray}
The vertex correction in Fig.~\ref{low} is this type.

\noindent{\bf 2)} Similarly, let's consider 
the case of $p_1^2=p_2^2=0$ and $t=(p_1-p_2)^2=p_3^2$.

In this case,
\begin{eqnarray}
&& C_1 = -{B_0(0)\over t} + {B_0(t)\over t}, \nonumber \\
&& C_2 = -{B_0(0)\over t} + {B_0(t)\over t},\nonumber \\
&& C_{00} = {1\over 4} + {B_0(t)\over 4},\nonumber \\
&& C_{11} = {B_0(0)\over 2 t} - {B_0(t)\over 2 t},\nonumber \\
&& C_{22} = {B_0(0)\over 2 t} - {B_0(t)\over 2 t},\nonumber \\
&& C_{12} = {1\over 2 t}.\nonumber \\
\end{eqnarray}

The vertex correction is
\begin{eqnarray}
V^\mu&=&(2-n)\left\{\left((2-n) C_{00} + tC_{12}\right)\gamma^\mu
+2\!\!\not{p_2}p_2^\mu C_{22}+2\!\!\not{p_2}p_1^\mu C_{12}\right\}
\nonumber\\
&&+(2-n)2\!\!\not{p_2}p_2^\mu C_2 -2\left(2\!\!\not{p_2}p_1^\mu + t \gamma^\mu \right) C_1.
\end{eqnarray}
We get
\begin{eqnarray}
V^\mu&=&\gamma^\mu\left(\frac{2}{\omega}-4+L_t\right)
+\frac{2\!\not{p_2}}{t} p_1^\mu\left(\frac{2}{\omega}-5+2L_t\right)
+\frac{2\!\not{p_2}}{t} p_2^\mu\left(\frac{1}{\omega}-1+L_t\right),
\nonumber\\
&=&\gamma^\mu\left(\frac{2}{\omega}-4+L_t\right)
-\frac{6\!\not{p_3}}{t} p_1^\mu\left(\frac{1}{\omega}-2+L_t\right)
+\frac{2\!\not{p_3}}{t} p_3^\mu\left(\frac{1}{\omega}-1+L_t\right),
\end{eqnarray}
where $L_t=\ln(-t/\mu^2)$.
Vertices of diagrams $V_1$ and $V_6$ in Fig.~\ref{vfigs} are in this category.

\noindent{\bf 3)} $p_1^2=0$ $p_2^2=Q^2$ and $t=(p_1-p_2)^2=p_3^2$. 
$p_1\cdot p_2 = (Q^2-t)/2$ and 
${\not{p_2}}\!\!{\not{p}_1} =-{\not{p}_1}\!\!{\not{p_2}} +2 p_1\cdot p_2$
In this case the coefficients are
\begin{eqnarray}
C_1 &=& {B_0(0)\over (Q^2 - t)} - 2 Q^2 {B_0(Q^2)\over (Q^2 - t)^2} + 
   (Q^2 + t) {B_0(t)\over (Q^2 - t)^2} - Q^2 {C_0(0,t, Q^2)\over (Q^2 - t)},
\nonumber \\
C_2 &=& {B_0(Q^2)\over (Q^2 - t)} - {B_0(t)\over (Q^2 - t)},
\nonumber \\
C_{00} &=& {1\over 4} + Q^2{ B_0(Q^2)\over 4 (Q^2 - t)} - 
t {B_0(t)\over 4 (Q^2 - t)}, \nonumber \\
C_{11} &=& {Q^2\over (Q^2 - t)^2} - 
(3 Q^2 - t) {B_0(0)\over 2 (Q^2 - t)^2} + 
   3Q^2 {B_0(Q^2))\over (Q^2 - t)^3} 
\nonumber \\ 
&&- (3 Q^2 + 4 Q^2 t - t^2) {B_0(t)\over 2 (Q^2 - t)^3}
 + Q^2{ C_0(0,t, Q^2)\over (Q^2 - t)^2},
\nonumber \\
C_{22} &=& -{B_0(Q^2)\over 2 (Q^2 - t)} + {B_0(t)\over 2 (Q^2 - t)},
\nonumber \\
C_{12} &=& -{1\over 2 (Q^2 - t)} - Q^2{ B_0(Q^2)\over 2 (Q^2 - t)^2} + 
   Q^2 {B_0(t)\over 2 (Q^2 - t)^2}.
\end{eqnarray}
The vertex correction becomes
\begin{eqnarray}
V^\mu&=&(2-n)\left\{\left((2-n) C_{00} -Q^2 C_{22}- (Q^2-t)C_{12}\right)\gamma^\mu
+2\!\!\not{p_2}p_2^\mu C_{22}+2\!\!\not{p_2}p_1^\mu C_{12}\right\}
\nonumber\\
&&+(n-2)Q^2\gamma^\mu C_2 +(2-n)2\!\!\not{p_2}p_2^\mu C_2 
-4\!\!\not{p_2}p_1^\mu C_1 + 2(Q^2-t) \gamma^\mu  C_1
\nonumber\\
&=  &
\frac{\gamma^\mu}{(t-Q^2  )^2} \left(\frac{2}{\omega}
\left((L_t - L_{Q^2} -1)Q^4 +  (L_{Q^2} - L_t) Q^2 t +  t^2\right)+ (L_t -4)t^2
\right.\nonumber\\&&
     \left.\phantom{\frac{1}{1}}+
(4 - 3 L_{Q^2} - L_{Q^2}^2 + 2 L_t +
 L_t^2)Q^4 + (L_{Q^2}^2 - 3 L_{Q^2} + 3 L_t - L_t^2) Q^2 t  
\right)
\nonumber\\&&
+\frac{2p_1^\mu \not{p_2}}{(t-Q^2  )^3}\left(\frac{2}{\omega}
 \left(( L_t-1 -  L_{Q^2} ) Q^4 +  (L_{Q^2} -  L_t) Q^2 t +  t^2\right)+ (2 L_t - 5 )t^2
\right.\nonumber\\
&&\left.\phantom{\frac{1}{\omega}}
 + (
5 - 5 L_{Q^2} - L_{Q^2}^2 + 3 L_t + L_t^2 )Q^4  + (
  L_{Q^2}^2 - 5 L_{Q^2} + 5 L_t - L_t^2 )Q^2 t  
\right)
\nonumber\\&&
+\frac{2p_2^\mu \not{p_2}}{t-Q^2 }
 (L_t-L_{Q^2}  ),
\end{eqnarray}
where we have used the renormalization conditions for $B_0(0)$. 
Vertices of diagram $V_4$ and $V_9$ are in this category.

Here is a simple application of these methods in the case of 
$\sigma_{q\bar{q}\gamma^*}$ (at $q_T=0$). 
An explicit calculation gives to lowest order (in $d=4 - 2 \epsilon$ 
dimensions) (with $x= Q^2/s$)

\beq
\sigma^{0}_{q\bar{q}\gamma^*}(s_1,s_2)= {4\pi^2\alpha e_Q^2 \over 3 s}
\delta(1-x) 
\left[(1-h) -\epsilon(h -1) - \epsilon {s_1}_\perp\cdot {s_2}_\perp \right].
\label{lowest}
\eeq

Notice that, when we send ${s_1}_\perp$ and ${s_2}_\perp$ to zero, we reobtain the usual 
longitudinal helicity dependence of the matrix element. 
We observe that this cross section, away from $d=4$, does not conserve 
helicity and has a spurious dependence from transverse spin, 
due to the $\epsilon$-dependent terms in (\ref{lowest}). 

This violation originates from the t'Hooft-Veltman. 
Notice also that the cross section shows no dependence 
on the transverse spin at $d=4$. Therefore, any transverse spin scatters, 
while the longitudinal components of the spin four-vector have to satisfy an 
helicity selection rule (the (1-h) factor). 

As in the longitudinal case, away from 4 dimension the t'Hooft Veltman prescription seems to indicate that there is a transverse 
spin dependence to lowest order, which is clearly an artifact of the regularization. 

Using the methods of this appendix, 
one gets quite straightforwardly for the virtual corrections to the 
$\sigma_{q\bar{q}\gamma^*}$ cross section 
 
\beqa
 \sigma^V_{q\bar{q}\gamma^*}(s_1,s_2)&=& {4\pi^2\alpha e_Q^2 \over 3 s}\delta(1-z) 
\left[(1-h) -\epsilon(h -1) - \epsilon {s_1}_\perp \cdot {s_2}_\perp 
\right] \nonumber \\ 
&& \hspace{.5cm}\times {\alpha_s\over 2 \pi}C_F \left( {4\pi \mu^2\over Q^2}\right)^\epsilon {1\over \Gamma[1- \epsilon]}\left( -{2\over \epsilon^2} - {3\over \epsilon}
-8 + \pi^2\right).
\label{second}
\eeqa

\subsection{gluon-quark-quark vertex (type 2)}
\begin{eqnarray}
\!\!\!V^\mu\!=\!\frac{(2\pi\mu)^{4-n}}
{i\pi^2}\!\int \!\!d^n l \frac{\gamma^\nu 
(\!\not{l} + \!\!\not{p_1})\gamma^\lambda ((l+2p_2)^\nu g^{\lambda\mu}+
(l-p_2)^\lambda g^{\mu\nu} - (2l+p_2)^\mu 
g^{\nu\lambda})}{l^2(l+p_1)^2(l+p_2)^2}.
\end{eqnarray}
We get
\begin{eqnarray}
V^\mu &=  & 2B_0 \gamma^\mu+\!\!\not{p}_2\!\!\not{p}_1\gamma^\mu( 2C_1 +C_2 +2C_0)
 - \gamma^\mu \!\!\not{p}_1\!\!\not{p}_2 (C_1-C_2+C_0) +\gamma^\mu p_2^2 C_2
\nonumber\\&&+ (n-2)\!\!\not{p}_2
(p_2^\mu (C_2 +2C_{22}) +2p_1^\mu C_{12}) + 2(n-2) \gamma^\mu C_{00}. 
\end{eqnarray}

\noindent{\bf 1)} In the case $p_1^2=0, p_2^2=Q^2$ and $(p_1-p_2)^2=0$ 
we get
\begin{eqnarray}
V^\mu= 2B_0 \gamma^\mu+2Q^2\gamma^\mu( C_1 +C_2 +C_0) 
 + 2(n-2) \gamma^\mu C_{00}
\end{eqnarray}
where
\begin{eqnarray}
B_0 &= & B_0(0), \\
C_0 &= & C_0(0,0,Q^2) .
\end{eqnarray}
The vertex correction is
\begin{equation}
V^\mu=\gamma^\mu \left(\frac{4}{\omega} -2 +L_{Q^2}\right) .
\end{equation}

\noindent{\bf 2)} $p_1^2=p_2^2=0$ and $t=(p_1-p_2)^2=p_3^2$. 

\begin{eqnarray}
V^\mu &=  & 2B_1 \gamma^\mu-  t\gamma^\mu( 2C_1 +C_2 +2C_0) - 2
p_1^\mu\not{p}_2 (C_1-C_2+C_0) 
\nonumber\\&&+ (n-2)\!\!\not{p}_2
(p_2^\mu (C_2 +2C_{22}) +2 p_1^\mu C_{12}) + 2(n-2) \gamma^\mu C_{00}, 
\end{eqnarray}
where
\begin{eqnarray}
B_1 &=&  B_0(t) \nonumber\\
C_0 &= & C_0(0,0,t)
\end{eqnarray}
and 
\begin{equation}
C_2=-2C_{22}.
\end{equation}
Substituting these expressions and the renormalization conditions for $B_0(0)$
we get 
\begin{equation}
V^\mu= -\gamma^\mu \left(\frac{2}{\omega^2} + \frac{2L_t-3}{\omega} -
\frac{\pi^2}{6} +L_t^2\right) - 2\frac{\not{p}_2}{t}p_1^\mu
\left(\frac{1}{\omega^2} + \frac{L_t}{\omega} -1 -\frac{\pi^2}{12}
+\frac{L_t^2}{2}\right).
\end{equation}
This vertex correction appears in the diagrams $V_3$ and $V_8$ of Fig.~\ref{vfigs}.

Notice that in all of the expressions given above, all the logarithmic 
functions containing a functional dependence in $Q^2$ or $s$ of the form 
$\ln (-Q^2/\mu^2)$ or  $\ln (-s/\mu^2)$ 
should be analytically continued as $\ln (Q^2/\mu^2) + i \pi$ and 
$\ln ( s/\mu^2) + i \pi$ 
respectively. Notice that due to cancelations of the imaginary parts 
because of interference, we usually get $\ln^2 (-Q^2/\mu^2)\to 
\ln^2 (Q^2/\mu^2) - \pi^2$ and similar ones with $Q^2$ replaced by $s$.

\section{Box contributions and tensor reductions}
\subsection{massless box diagram}
The diagram is evaluated in the Euclidean region ($s<0$, $t>0$) and analytically continued to the physical region using the typical relations 
\begin{eqnarray}
&& (-s)^{-\epsilon}\to |s|^\epsilon e^{i\pi \Theta(s)}\nonumber \\
&& \ln (-s)\to \ln|s|-i \pi \Theta(s)
\end{eqnarray}
where $\Theta(x)$ is the step function. There are various expressions of the 
massless scalar box, with/without Spence functions which we list below. 
Their equivalence can be easily shown by using the relations 
\begin{eqnarray}
&&Li_2(x) + Li_2(1-x^{-1})= -{1\over 2} \ln^2(x),\,\,\,\,\,\, x>0\nonumber \\
&& \ln(1-x)=\ln(x-1) + i\pi.
\end{eqnarray}
\begin{eqnarray}
&& I_0\equiv\int {d^{4-2 \epsilon}p\over (2 \pi)^{4-2 \epsilon}p^2 (p- k_1)^2 
(p- k_1 - k_2)^2 (p- k_1 - k_2 - k_3)^2} \nonumber \\
&&= {i\over (4 \pi)^{2-\epsilon}}{c_\Gamma\over s t}\left[
{2\over \epsilon^2}\left( (-s)^\epsilon + (-t)^\epsilon\right)
- \ln^2(s/t) - \pi^2\right]\nonumber \\
&&={i\over (4 \pi)^2 s t}\left({-s\over 4 \pi}\right)\Gamma[1 + \epsilon]
\left( {4\over \epsilon^2} -{2\over \epsilon}\ln t/s - {5\over 3} \pi^2
\right) \nonumber \\
&&= {i (-4 \pi)^{(4-n)/2}\over 4\pi^2}\Gamma\left[{4-n\over 2}\right]
B\left[{n\over 2}-1,{n\over 2}-1\right] 2 (n-3) A_1
\end{eqnarray}

where 
\begin{eqnarray}
&& c_\Gamma= {\Gamma[1 + \epsilon]\Gamma^2(1-\epsilon)\over 
\Gamma(1- 2\epsilon)} \nonumber \\
&& A_1= {1\over s t}\left({1\over \epsilon}( s^{-\epsilon} + t^{-\epsilon})
 -  s^{-\epsilon}(1 + {s\over t}) J( 1 + {s\over t}) - t^{-\epsilon}
(1 + {t\over s})J(1 + {t \over s})\right)
\nonumber \\
&& J(x)=-{\ln (1-x)\over x} +{\epsilon\over x}Li_2(x).
\end{eqnarray}
\subsection{massive box diagram (scalar)}
The expression of the direct scalar box diagram with one external mass 
has been given in \cite{vn} (here $n=4 -2\epsilon$)
\begin{eqnarray}
&& I_4(s,t, Q^2)=-i (4\pi)^{-n/2}{2\over \epsilon^2}c_\Gamma 
\left( {(-Q^2)^{-\epsilon}\over s t}F(1,-\epsilon,1-\epsilon,-{u Q^2\over s t})
\right.\nonumber \\
&& \,\,\,\,\,\,\,\left. -{(-s)^{-\epsilon}\over s t}F(1,-\epsilon,1-\epsilon, 
-{u\over t}) -{(-t)^{-\epsilon}\over s t}F(1,-\epsilon,1-\epsilon, -{u\over s})
\right)
\label{van}
\end{eqnarray}
 This expression can be cast into a simpler form 
by some manipulations which we are going to discuss here briefly.
We use the expansion
\begin{eqnarray}
&& F(1,-\epsilon,1-\epsilon, z)= 1 + \epsilon \ln(1-z) + \epsilon^2 S(-z)
\nonumber \\
&& S(z)\equiv \int_0^z dx{\ln(1+x)\over x}=-Li_2(-z).
\end{eqnarray}
$S(z)$ is the Spence function, related to the Euler dilogarithm. 
Expanding in $\epsilon$ Eq.~(\ref{van}) we get 
\begin{eqnarray}
 I_4(s,t, Q^2)=-i (4\pi)^{-n/2}c_\Gamma\left(
{2\over \epsilon^2}\left[ (-Q^2)^{-\epsilon} -(-s)^{-\epsilon}
- (-t)^{-\epsilon}\right] + 2 \log_1 + 2 \Sigma\right),
\end{eqnarray}
where 
\begin{eqnarray}
&& \log_1\equiv -\ln(-Q^2)\ln( 1+ {u Q^2\over s t}) + \ln(-s)\ln(1 + {u\over t})
+ \ln(-t)\ln(1+{u\over s})\nonumber \\
&&\Sigma\equiv S\left({u Q^2\over s t}\right) -S\left({u\over t}\right)
- S\left({u\over s}\right).
\end{eqnarray}
We use  the relations 
\begin{eqnarray}
&& Li_2(z) + Li_2(1-z)=-\ln(1-z)\ln z + {\pi^2\over 6},\nonumber \\
&& Li_2(x y)= Li_2(x) + Li_2(y) - Li_2\left( {x (1-y)\over 1- x y}\right)
- Li_2\left( {y(1-x)\over 1 - x y}\right) \nonumber \\
&&\,\,\,\,\,\,- \ln \left({1-x\over 1 - x y}\right)\ln\left({1- y\over 1- x y}\right),
\nonumber \\
&& \left(1 + {u Q^2\over s t}\right)= \left(1 - {Q^2\over s}\right)\left(1- {Q^2\over t}\right)= 
\left(1 +{u\over s}\right)\left(1 +{u\over t}\right),\nonumber \\
&& S(x)=-S\left({1\over x}\right) + {\pi^2\over 6} + {1\over 2 }\ln^2 x
\end{eqnarray}
to relate dilogs of different arguments in order to get
\begin{equation}
\Sigma=Li_2\left(1- {Q^2\over s}\right) +Li_2\left(1- {Q^2\over t}\right)
+ \log_2 - {5\over 6}\pi^2,
\end{equation}
and
\begin{eqnarray}
\log_2&=&\ln\left( 1 + {u Q^2\over s t}\right)\ln\left({-u Q^2\over s t}\right)
-{1\over 2}\ln^2 \left(\frac{u}{t}\right) -{1\over 2}\ln^2\left({u\over s}\right) 
\nonumber \\
&&\hspace{-1cm}+ \ln \left(1 + {t\over u}\right) 
\ln\left(-{u\over t}\right)
 + \ln \left(1 + {s\over u}\right)\ln \left(-{s\over u}\right) 
- \ln\left(-\frac{s}{u}\right)\ln\left(-\frac{t}{u}\right).
\end{eqnarray}
After analytic continuation we get
\begin{equation}
\log_1 +\log_2= \frac{1}{2}\ln^2\left({|s|\over |t|}\right) +{\pi^2\over 2}
\end{equation}
and the final expression 
\begin{eqnarray}
&& I_4(s, t, Q^2)=i {(4\pi)^{-n/2}\over s t}c_\Gamma\ 
\left[
{2\over \epsilon^2}\left[- (-Q^2)^{-\epsilon} +(-s)^{-\epsilon}
+ (-t)^{-\epsilon}\right]\right.\nonumber \\
&&\,\,\,\,\,\,\,\,\,\,\,\,\left. -2 Li_2\left(1 - {Q^2\over s}\right) 
-2 Li_2\left(1 - {Q^2\over t}\right)
- \ln^2\left({|s|\over |t|}\right) +{2\pi^2\over 3}\right].
\label{bbox}
\end{eqnarray}
Notice that this relation needs analytic continuation in order to be applied 
to the calculation of the scalar box diagram. Its range of validity is in the 
unphysical region $Q^2<0$, $s<0$ and $t<0$, therefore 
we analytic continue it to the region $Q^2>0$, $s>0$ and $t<0$ with the relations

\begin{eqnarray}
&& (-s)^{-\epsilon}\to |s|^\epsilon e^{i\pi}\nonumber \\
&& (-Q^2)^{-\epsilon}\to |t|^\epsilon e^{i\pi}\nonumber \\
\end{eqnarray}
and leave $t$ unchanged. Notice that there are no imaginary parts generated by this continuation, as one can easily check from (\ref{bbox}).

In the case of the scalar $u$-channel diagram $V_{11}$ we replace $s\to u$ 
in (\ref{bbox})
and therefore we analytically continue only in $Q^2$. There are imaginary parts 
generated by this procedure, which in this case don't cancel. However, since the interference terms are of the form $A_{box}\times B^*_{Born} + A^*_{box}\times B_{Born}$ in the differential 
cross section, than we get again a cancelation of the imaginary parts. 
The structure of the di-logs contribution generated by (\ref{bbox})
is different from ref.~\cite{EMP}. We have used 
\begin{eqnarray}
Li_2\left({Q^2\over Q^2 - t}\right)&=& - Li_2\left(1- {Q^2\over t}\right)- \ln\left(1- {Q^2\over t}\right)
\ln\left( {Q^2\over t}\right) + {\pi^2\over 6} \nonumber\\
&&+ {1\over 2}\ln^2\left({Q^2\over t}\right) 
-{1\over 2}\ln^2\left({-Q^2\over Q^2 - t}\right)\nonumber \\
 Li_2\left({Q^2\over s}\right)&=& - Li_2\left(1- {Q^2\over s}\right) - \ln\left( {Q^2\over s}\right)
\ln\left({Q^2\over s}\right) + {\pi^2\over 6} 
\end{eqnarray}
to further simplify the result for the virtual corrections.

\newpage
\section*{Figures}
\vfill
\vfill
\begin{figure}
\centerline{\epsfbox{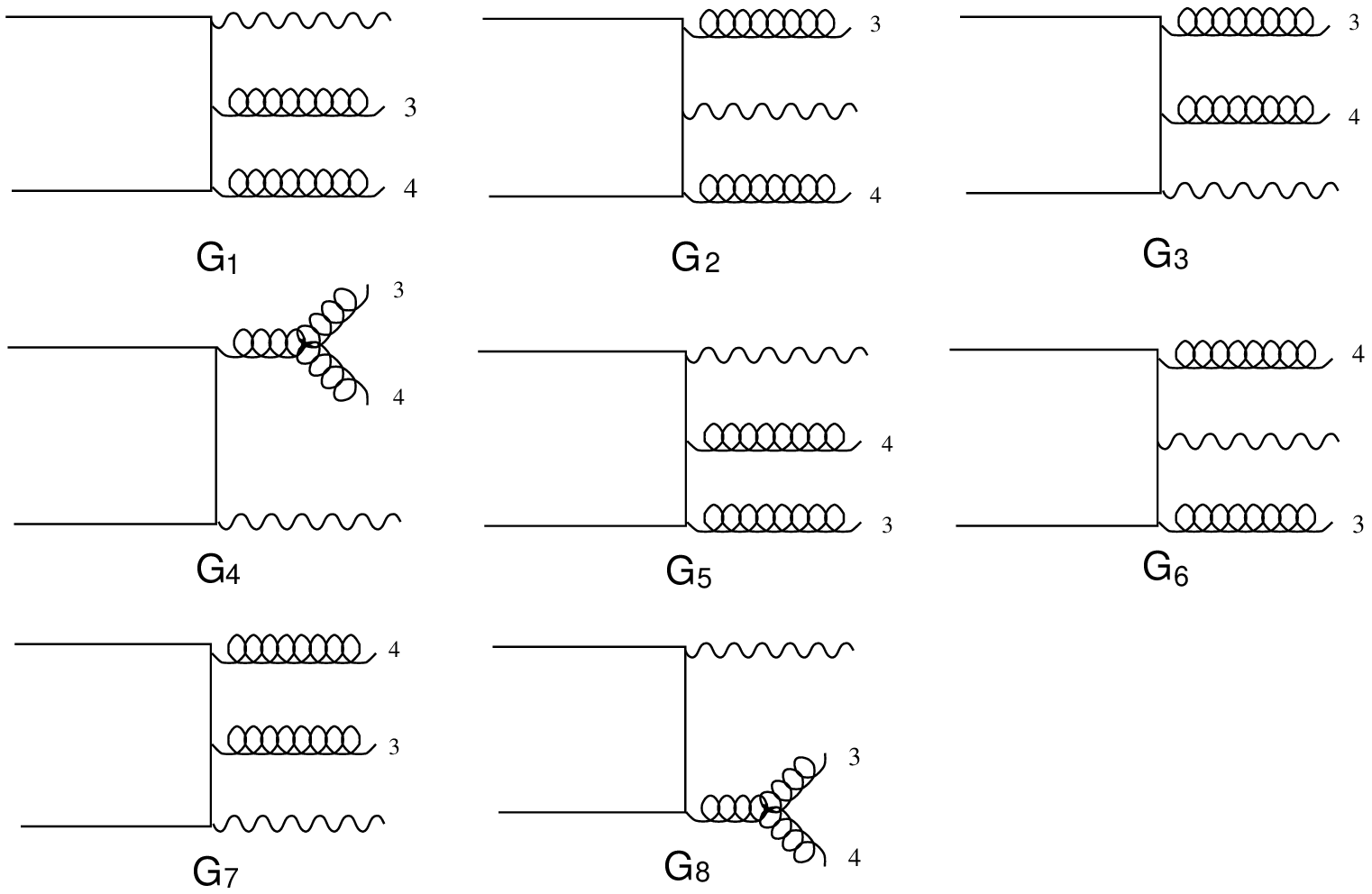}}
\caption{Diagrams which contribute to the process
$q+\bar{q}\rightarrow \gamma^* +G+G$}
\end{figure}
\begin{figure}
\centerline{\epsfbox{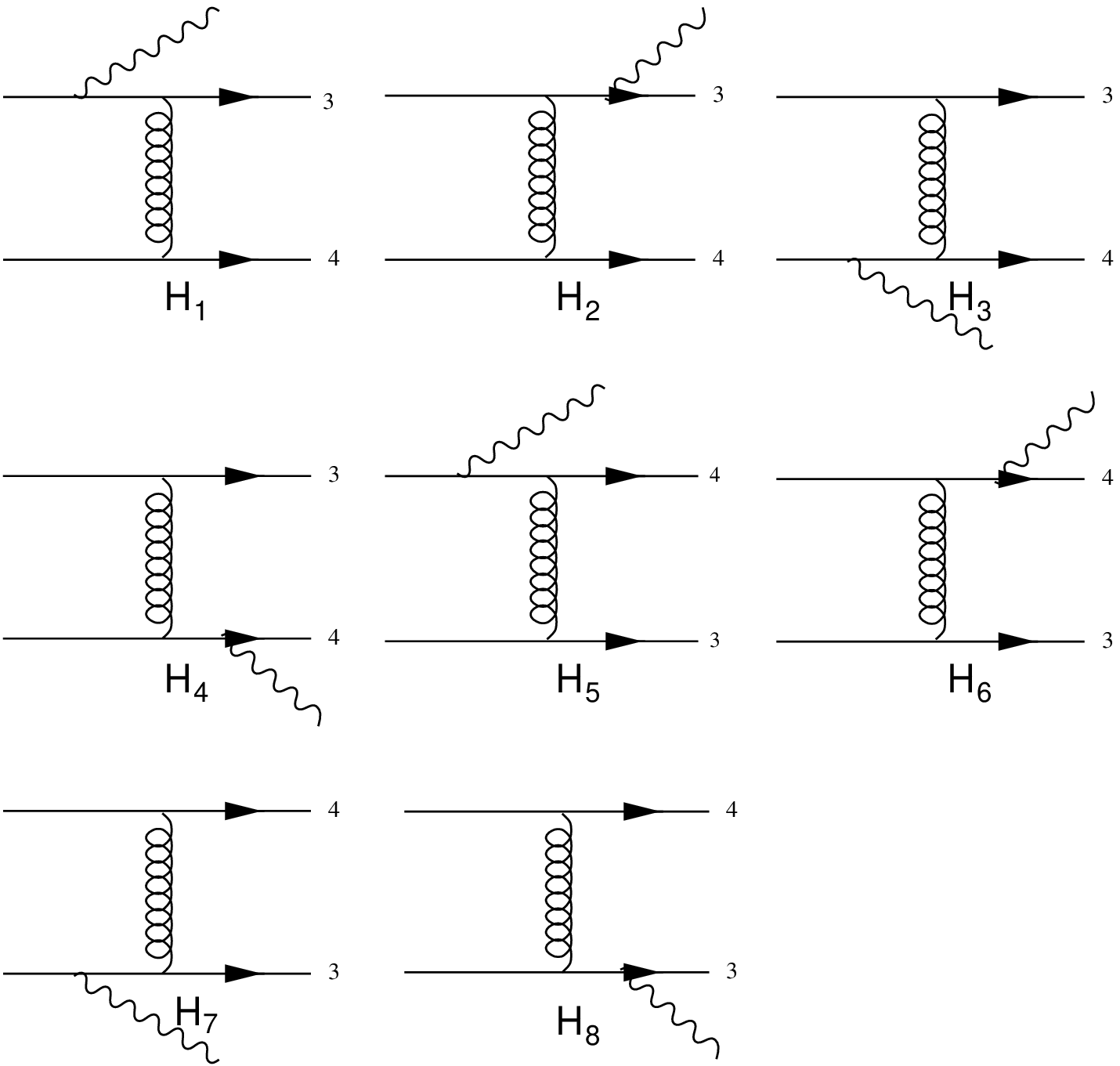}}
\caption{Diagrams which contribute to the process
$q+q\rightarrow \gamma^* +q+q$}
\end{figure}
\begin{figure}
\centerline{\epsfbox{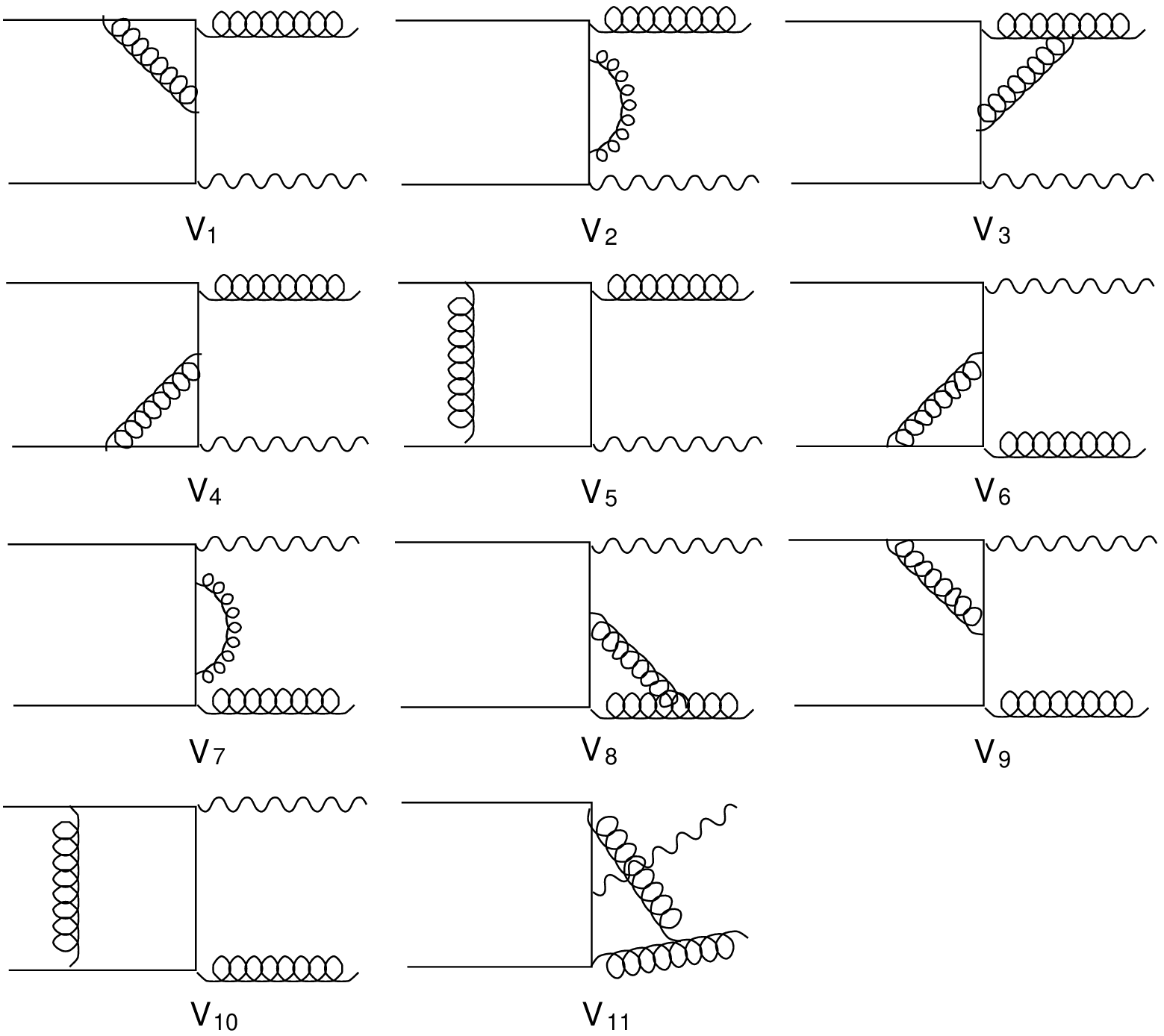}}
\caption{Diagrams which contribute to the process
$q+\bar{q}\rightarrow \gamma^* +G$}
\label{vfigs}
\end{figure}
%\begin{figure}
%\epsfxsize=140mm
%\centerline{\epsfbox{canc.eps}}
%\caption{Diagrams involved in the cancelation of the $s_2 \rightarrow 0 $
%singularity}
%\label{canc}
%\end{figure}

\end{document}